\pdfoutput=1
\documentclass[twoside,english,preprint,3p]{elsarticle}
\usepackage[T1]{fontenc}
\usepackage[latin9]{inputenc}
\usepackage{babel}
\usepackage{verbatim}
\usepackage{prettyref}
\usepackage{float}
\usepackage{url}
\usepackage{amsmath}
\usepackage{amssymb}
\usepackage{graphicx}
\usepackage{esint}
\usepackage[unicode=true,
 bookmarks=false,
 breaklinks=false,pdfborder={0 0 1},backref=false,colorlinks=false]
 {hyperref}

\makeatletter

\newcommand{\lyxdot}{.}

\floatstyle{ruled}
\newfloat{algorithm}{tbp}{loa}
\providecommand{\algorithmname}{Algorithm}
\floatname{algorithm}{\protect\algorithmname}

\journal{arXiv}

\PassOptionsToPackage{utf8}{inputenc}
\usepackage{inputenc}
\usepackage{pgfplots}

\usepackage{import}
\usepackage{epstopdf}

\usepackage[font=footnotesize,labelfont=bf]{caption}
\usepackage{subfigure}

\makeatother

\begin{document}

\title{Adaptive and Iterative Methods for Simulations of Nanopores with
the PNP-Stokes Equations}

\author[tuw]{Gregor~Mitscha-Baude\corref{cor1}}

\ead{gregor.mitscha-baude@gmx.at}

\author[tuw]{Andreas~Buttinger-Kreuzhuber}

\author[tuw]{Gerhard~Tulzer}

\author[tuw]{Clemens~Heitzinger}

\cortext[cor1]{Corresponding author}

\address[tuw]{TU Wien, Wiedner Hauptstrasse 8--10, A-1040 Vienna, Austria}
\begin{abstract}
We present a 3D finite element solver for the nonlinear Poisson-Nernst-Planck
(PNP) equations for electrodiffusion, coupled to the Stokes system
of fluid dynamics. The model serves as a building block for the simulation
of macromolecule dynamics inside nanopore sensors.

We add to existing numerical approaches by deploying goal-oriented
adaptive mesh refinement. To reduce the computation overhead of mesh
adaptivity, our error estimator uses the much cheaper Poisson-Boltzmann
equation as a simplified model, which is justified on heuristic grounds
but shown to work well in practice. To address the nonlinearity in
the full PNP-Stokes system, three different linearization schemes
are proposed and investigated, with two segregated iterative approaches
both outperforming a naive application of Newton's method. Numerical
experiments are reported on a real-world nanopore sensor geometry.

We also investigate two different models for the interaction of target
molecules with the nanopore sensor through the PNP-Stokes equations.
In one model, the molecule is of finite size and is explicitly built
into the geometry; while in the other, the molecule is located at
a single point and only modeled implicitly -- after solution of the
system -- which is computationally favorable. We compare the resulting
force profiles of the electric and velocity fields acting on the molecule,
and conclude that the point-size model fails to capture important
physical effects such as the dependence of charge selectivity of the
sensor on the molecule radius.

\end{abstract}
\maketitle

\section{Introduction}

\begin{figure}
\begin{centering}
\includegraphics{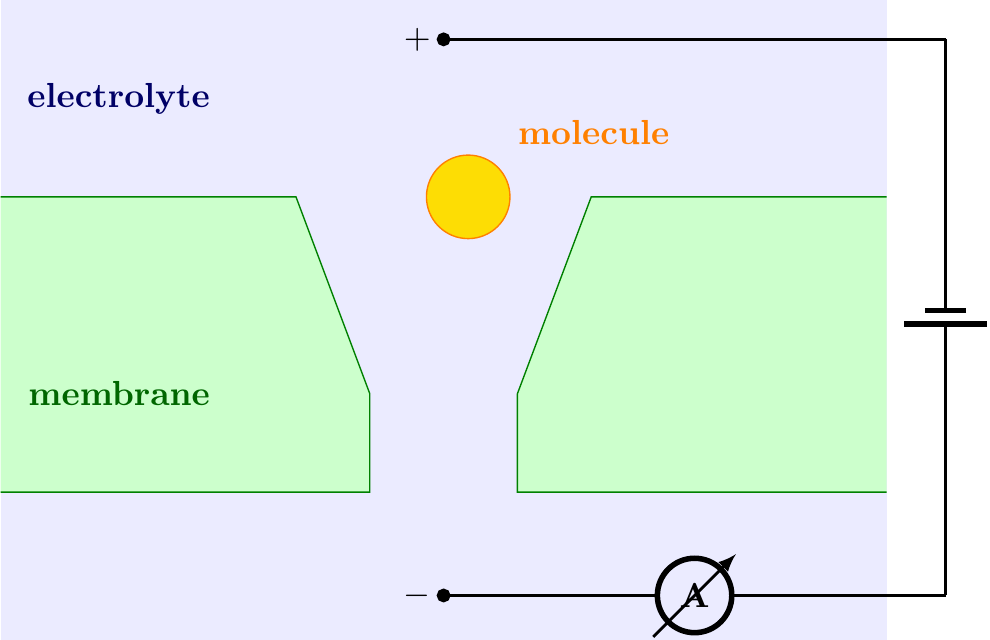}
\par\end{centering}

\caption{\label{fig:setup}Schematic of a nanopore sensor. The pore connects
two large electrolyte reservoirs (blue) separated by a membrane (green).
A transmembrane voltage is applied (illustrated by two electrodes
in a circuit diagram), triggering an ionic current through the pore.
Target molecules (yellow) can be detected based on modulation of the
current as they enter the nanopore.}
\end{figure}

Nanopore sensors are biotechnological devices designed to mimic the
functionality of ion channels that occur in organic cells. They have
shown promise as a tool to detect and analyze single molecules in
an electrolyte solution. This has enabled fast and cheap DNA sequencing
\citep{Bayley2015,Schneider2012} -- recently put to practice for
surveillance of the Ebola virus in West Africa \citep{Quick2016}
--, and is also finding applications in protein detection and sequencing
\citep{Howorka2012}%
. The basic principle of a nanopore sensor is illustrated in Figure~\ref{fig:setup}.
Since nanopore technology is in an experimental state, researchers
need to be able to investigate new sensor designs rapidly, and simulations
play an important role in this process.

The main continuum model for nanopores are the steady-state Poisson-Nernst-Planck
(PNP) equations \citep{coalson-kurnikowa2005-pnp,Kuyucak2001,Maffeo2012},
which capture the electrodiffusion of various ion species in solution.
They can be augmented by the Navier-Stokes equation to include effects
of electroosmotic flow, resulting in the \emph{Poisson-Nernst-Planck-Stokes
(PNPS)} equations, which will be the focus of our work. Pioneered
by \citet{rubinstein1990electro}, they have been used to model, for
example, lab-on-chip devices \citep{Erickson2002a,Erickson2002b},
biological ion channels \citep{Wei2012a,Chen1995} and solid-state
nanopores \citep{keyser09origin,Keyser2010,Ghosal2006}.

Given the importance of the PNPS model in engineering, relatively
few papers have addressed the numerical approach in detail. Splitting
schemes for the time-dependent formulation have been analysed in \citep{Prohl2010,Sacco2015,Metti2016};
but our interest lies in the steady-state distributions which can
be obtained more efficiently directly, without time-stepping. Most
works related to nanopore modeling rely on a black-box implementation
provided by the commercial Comsol Multiphysics package \citep{Vlassiouk2008,Vlassiouk2008b,Lu2012,Laohakunakorn2015,Mao2013}.
In the microfluidic community, special-purpose finite difference codes
have been shown to give better performance \citep{Karatay2015}. For
our work, however, we prefer the flexibility of unstructured meshes
to account for complex real-world nanopore geometries, which are defined,
for instance, by the shapes of biological channel proteines like $\alpha$-hemolysin
\citep{Song1996a}. Therefore, finite elements will be used for discretization.

Our goal in this work was to create a state-of-the-art PNPS solver
that overcomes the limitations of both off-the-shelf commercial solutions
and highly structured geometries. The two main contributions to the
numerical literature are
\begin{itemize}
\item a detailed investigation of linearization schemes for PNPS, and
\item a fast adaptive mesh refinement scheme based on goal-oriented error
estimation.
\end{itemize}

The adaptivity part is specifically tailored towards fast and reliable
evaluation of the electrophoretic force on particles surrounded by
the fluid, which we consider important for nanopore sensor modeling.
The novelty compared to previous applications of goal-oriented adaptivity
\citep{Becker2003,bangerth2013adaptive}, besides being the first
for PNPS, is that we base the estimator not on the full PDE system
but on a simplified model inspired by physics, namely, the linear
Poisson-Boltzmann equation. Since solving the latter equation takes
only a fraction of the time compared to the full system, this essentially
renders the whole iterative mesh refinement procedure a cheap preprocessing
step to be performed before the actual simulation. While by computing
the estimator for a different equation we necessarily leave the realm
of theoretical validity (where sometimes convergence of adaptive methods
can be proved rigorously \citep{feischl2016abstract,Feischl2014,Carstensen2014}),
numerical results indicate that our approach does work surprisingly
well (Section \ref{sub:Assessment-of-goal-adaptive}). The estimator
not only qualitatively picks the correct regions for refinement, but
also drives down the error in quantities of interest of the full model
with the theoretically optimal rate $O(h^{2})$. 

Our findings on linearization are of general interest for steady-state
PNPS related models. To solve the PNPS equations -- which form a nonlinear
system of seven coupled equations --, we consider three linearization
schemes: a monolithic Newton method applied to the whole system, a
Gummel-type fixed-point method, and a hybrid scheme where Newton's
method is only applied to the PNP part. The fixed-point method uses
a novel corrected Poisson equation which lowers iteration count significantly
compared to similar approaches in the literature \citep{Lu2007}.%
{} In our numerical comparisons, the fixed-point and hybrid methods
both reveal some strengths and weaknesses, but clearly outperform
the Newton method.

A third contribution of this work lies in modeling and is more specifically
tied to the simulation of nanopore sensors. For sensor prototyping,
solution of the PNPS system is just one building block to obtain the
force which drives the transport of target molecules through the sensor.
We investigate two different models for the force where the molecule
is either finite-sized or point-sized. In the point-size model, the
molecule has no influence on its surrounding electric field and fluid
flow; this is a strong simplification, since the force on the whole
domain can be obtained from a single PNPS solve. Indeed, we find that
the two models deviate significantly from each other. However, we
show how to calibrate the point-size model using a single evaluation
of the finite-size model, to obtain a much better fit.

The remainder of the paper is organized as follows. The PNPS equations,
including a description of boundary conditions and other modeling
aspects are presented in Section \ref{sec:Model}. We also introduce
a 2D version of the system in cylindrical coordinates that can simplify
the simulation whenever the geometry is axisymmetric. Section \ref{sec:Numerical-Method}
is devoted to numerical methods. Linearization of PNPS is discussed
in \ref{sub:Linearization-of-PNPS}, and further details of the numerical
solver are given in Section \ref{sub:Details-of-the-numerical}. In
Section \ref{sub:Goal-oriented-adaptivity} the goal-oriented adaptivity
framework is introduced. We formulate an adaptive algorithm similar
to \citep{Becker2003,Rognes2012} which makes use of patch-wise extrapolation
of the dual solution. Additionally, we propose a cheaper variant without
extrapolation where the error estimator is based directly on the dual
solution.

Numerical results are presented in Section \ref{sec:Results-and-Discussion}.
The geometry we use for experiments is modeled after the DNA-based
nanopore sensor recently published in \citep{Burns2016}, where we
carefully try to replicate the experimental set-up. In the Appendix,
additional numerical results can be found which validate our solver
against the exact solution on an idealized test problem. We conclude
with Section \ref{sec:Conclusion}.

\section{Model\label{sec:Model}}

\global\long\def\u{\boldsymbol{u}}
\global\long\def\grad{\nabla}
\global\long\def\Div{\grad\cdot}
\global\long\def\laplace{\Delta}
\global\long\def\F{\boldsymbol{F}}
\global\long\def\x{\boldsymbol{x}}
\global\long\def\v{\boldsymbol{v}}
\global\long\def\j{\boldsymbol{j}}
\global\long\def\R{\mathbb{R}}
\global\long\def\n#1{\mathbb{\|}#1\|}

To start with a broad picture of what simulations of a nanopore sensor
should achieve, consider the target molecule in Figure \ref{fig:setup},
and suppose the sensor is designed to detect exactly this molecule,
which is of a certain shape and carries a certain charge. Two basic
questions regarding the functionality of our sensor would be, first,
whether the molecule can even enter the nanopore, and second, whether
it would translocate the pore sufficiently slowly to be detected by
measurements of ionic current. Both questions can for instance be
addressed using the framework of the Langevin equation \citep{langevin1908theorie}
for the dynamics of small particles.

In this or any similar framework, the specifics of the physical environment
(i.e., sensor and molecule) enter through the electrophoretic force
$\F(\x)$ on the target molecule at any given position $\x$. Once
the force is available in parametrized form, we can devise algorithms
to compute quantities of interest such as mean translocation time
and probability. The physical model underlying $\F(\x)$ is encapsulated
in the PNPS equations, which are the focus of our work and are discussed
now.

\subsection{PNPS equations}

The steady-state Nernst-Planck (NP) equations for a 1:1 electrolyte
with positive and negative ion concentrations $c^{+}$ and $c^{-}$,
respectively, are given by\begin{subequations}\label{eq:nernst-planck}
\begin{align}
\Div\j^{\pm} & =0,\\
\j^{\pm} & =-D^{\pm}\grad c^{\pm}\mp\frac{qD^{\pm}}{kT}c^{\pm}\grad\phi+c^{\pm}\u.
\end{align}
\end{subequations}The quantity $\j^{\pm}$ is called the ionic flux;
$D^{\pm}$ is the ion-specific diffusion coefficient; $q$ the elementary
charge; $\phi$ the electric potential; and $\u$ the fluid velocity.
This differs from the more well-known form of the Nernst-Planck equations
\citep{coalson-kurnikowa2005-pnp} by the convective flux term $c^{\pm}\u$,
which introduces coupling to the Stokes equation; the term describes
transport of ions with the surrounding background medium, water. The
ion concentrations $c^{\pm}$ are measured in molar units $\text{mol}/\text{m}^{3}$
and will therefore be multiplied by the Faraday constant $F=96485\text{ C}/\text{mol}$
to give rise to a charge concentration.

Notably, we will not assume that the diffusion coefficients $D^{\pm}$
are constant and equal to their bulk value. It has been recognized
in the literature that the confining environment of a narrow channel
can substantially reduce diffusivity \citep{Noskov2004,Paine1975,Egwolf2010,aksimentiev12-BD,Heitzinger2011,Khodadadian2015},
so we have built this possibility into our solver. In the numerical
experiments below, we use a bulk value of $D^{\pm}=1.9\text{ nm}^{2}/\text{ns}$
for both ions, which is lowered by a factor of $0.5$ inside nanopores.
For absolute temperature we use $T=293$ K.

The Poisson equation for the electric potential $\phi$ is
\begin{equation}
-\nabla\cdot\left(\varepsilon\nabla\phi\right)=\rho_{0}+F(c^{+}-c^{-})\label{eq:poisson}
\end{equation}
with $\varepsilon$ denoting the material-dependent electric permittivity
and $\rho_{0}$ permanent charges present in the system, e.g. in target
molecules and on the pore walls. Equivalently, permanent charges located
on a surface are incorporated via Neumann boundary or interface conditions.

Finally, the fluid velocity field $\u$ and pressure $p$ are determined
by the Stokes equations, which read, in conservative form,\begin{subequations}\label{eq:stokes}
\begin{eqnarray}
-\Div\left[\eta\left(\grad\u+\grad\u{}^{T}\right)-pI\right] & = & -F(c^{+}-c^{-})\grad\phi,\\
\Div\u & = & 0.
\end{eqnarray}
\end{subequations}Here, $\eta=10^{-3}$ $\text{Pa}\cdot\text{s}$
is the viscosity and $I$ is the identity tensor. The right hand side
is the force arising from ions which are pushed by the electric field,
dragging along nearby water molecules. The nonlinear inertial term
$\u\cdot\grad\u$ usually present in the Navier-Stokes equations can
be neglected, since the Reynolds number in nanopore-related systems
is very low%
\footnote{For the typical length scale of about $L=10^{-8}$ m, velocity of
$u=10^{-2}$~$\text{m}/\text{s}$ and fluid density of $\rho=10^{3}$
$\text{kg}/\text{m}^{3}$, we have 
\[
\textrm{Re}=\frac{\rho uL}{\eta}=10^{-4}.
\]
}.

Taking together \eqref{eq:nernst-planck}--\eqref{eq:stokes}, the
PNPS system consists of seven equations for the seven unknowns: the
electric potential $\phi$, the two ion concentrations $c^{+}$ and
$c^{-}$, the three components of the velocity $\u$ and the pressure
$p$.

\subsection{Geometry and boundary conditions}

We solve PNPS on a space domain like that pictured in \prettyref{fig:setup}.
It consists of an aqueous pore embedded in a solid membrane and possibly
a solid sphere inside or near the pore which represents the target
molecule. The membrane separates two cylindrical reservoirs of electrolyte
which extend the computational domain to about $20$ nm in every direction;
this has been empirically confirmed large enough to ensure that the
physics inside the pore are not disturbed by artificial boundary conditions,
by comparing with reservoir sizes of up to $1000$ nm.

The Stokes and Nernst-Planck equations for the velocity and ion concentrations
are solved only in the fluid part of the domain; the Poisson equation,
on the other hand, is solved on the whole domain including membrane
and molecule. The pore/membrane may consist of several distinct dielectric
materials depending on the exact sensor to be modeled. Details for
a concrete application set-up are provided in Section \ref{sub:Model-of-DNA-based}.

Boundary conditions are specified as follows. For the ions, we fix
concentrations on the top and bottom of the computational domain at
a bulk value $c^{\pm}=c_{{\rm 0}}$ (Dirichlet condition), while on
other boundaries we apply a no-flux (Neumann) condition, $n\cdot\j^{\pm}=0$,
which models hard repulsion at the pore walls.

The external electric field is applied by setting the potential to
$\phi=0$ on the top and to a suitable bias such as $\phi=-0.1$ V
on the bottom. On the outer barrel of the cylindrical reservoirs,
the Neumann condition $\partial_{n}\phi=0$ is applied, assuming that
sufficiently far from the influence of the pore the potential varies
only in the direction perpendicular to the membrane. Interface conditions
are used to incorporate charges sitting on the pore walls, i.e. $[\varepsilon\partial_{n}\phi]=\rho$
where $[\,\cdot\,]$ denotes the jump across the interface and $\rho$
the surface charge density. Charges on the target molecule, on the
other hand, are incorporated via a volume term $\rho_{0}$ on the
right hand side of the Poisson equation \eqref{eq:poisson}; in this
way, the electric force on the molecule can be evaluated consistently
by integrating $-\rho_{0}\grad\phi$ over the molecule.

For the fluid velocity, we use the no-slip condition $\u=0$ on fluid-solid
interfaces and the natural stress-free boundary condition on the outer
reservoir boundaries. In addition, we fix the pressure by setting
it to zero on the reservoir top.

\subsection{Axisymmetric reduction}

Most nanopores can be modeled to a good approximation as symmetric
around the central $z$ axis. In fact, all the 3D geometries we implemented
can be generated by rotation of a two-dimensional cross-section about
this axis. In such a setting, if the target molecule is either not
present at all or sits on the central axis, all the equations will
possess axial symmetry. By transforming to cylindrical coordinates
$r,\ \varphi,\ z$ and using the fact that the solutions do not depend
on the angular component $\varphi$, the PNPS equations become a system
of PDEs in the two variables $r,\ z$. This 2D axisymmetric version
is much faster to solve and was implemented in addition to the full
3D version to provide validation and rapid calculations for set-ups
without off-centered molecules.

The idea of using axial symmetry to reduce computational effort is
not new in the context of nanopores. In fact, \citep{Lu2012,Laohakunakorn2015,Mao2013,Mao2014}
-- which are the publications most similar in scope to this one --
all restrict themselves to axisymmetric settings, so the novelty of
our implementation rather lies in the fact that we handle the full
3D case as well.

\subsection{Physical Quantities of Interest\label{sub:Physical-Quantities}}

Since we want to describe translocation of target molecules through
a nanopore, the primary output of our PNPS solver will be the mean
force $\F$ acting on such a molecule, which we model as a sum of
two contributions 
\begin{equation}
\F=\F_{{\rm el}}+\F_{{\rm drag}}.\label{eq:Ftotal}
\end{equation}
$\F_{{\rm el}}$ is the electric force induced by the electric field
acting on charges on the molecule, and $\F_{{\rm drag}}$ is the drag
force exerted on the molecule surface by the moving fluid.

\paragraph{Finite-sized molecule}

For our main model, we build a spherical target molecule into the
geometry by excluding its volume from the fluid domain and ensuring
that the mesh aligns with the molecule boundaries. Then, the forces
are given by
\begin{equation}
\F_{{\rm el}}=-\int_{M}\rho_{0}\grad\phi\label{eq:Fel}
\end{equation}
and
\begin{equation}
\F_{{\rm drag}}=\int_{\partial M}n\cdot\sigma\label{eq:Fdrag}
\end{equation}
respectively, where $M$ is the part of the domain occupied by the
molecule, $\rho_{0}$ the molecule surface charge density and $\sigma=\eta\left(\grad\u+\grad\u{}^{T}\right)-pI$
the fluid stress tensor.

Note that the drag force in formula \eqref{eq:Fdrag} is expressed
as a surface integral. In practice, we actually evaluate drag component-wise
as the volume integral
\begin{equation}
F_{{\rm drag}}^{i}=\int_{\Omega_{F}}\sigma:\grad\boldsymbol{z}_{i}-\boldsymbol{f}\cdot\boldsymbol{z}_{i}\label{eq:Fdrag-volume}
\end{equation}
where $\Omega_{F}$ is the fluid domain (excluding the molecule),
$\boldsymbol{f}=-F(c^{+}-c^{-})\grad\phi$ the volume force and $\boldsymbol{z}_{i}$
any vector field that satisfies $\boldsymbol{z}_{i}|_{\partial M}=\boldsymbol{e}_{i}$
where $\boldsymbol{e}_{i}$ is the $i$-th unit vector. This representation
is equivalent to \eqref{eq:Fdrag} if $\u$ and $p$ solve the Stokes
equation exactly, as can be shown via integration by parts. Formula
\eqref{eq:Fdrag-volume} was reported by others \citep{Becker2003}
and observed by ourselves to yield a more accurate numerical approximation.
For $\boldsymbol{z}_{i}$ we use a piece-wise linear interpolation
of $\boldsymbol{e}_{i}$ on the molecule boundary and zero elsewhere.

\paragraph{Point-sized molecule}

If the PNPS equations are solved on a geometry \emph{without} target
molecule, we can still obtain an estimate for $\F_{{\rm el}}$ and
$\F_{{\rm drag}}$ on every given position of the computational fluid
domain. In this case, we neglect the influence that the charge and
geometrical presence of the molecule has on the physical environment,
and assume the molecule to be located at a single point%
. The electric force is then given by
\begin{equation}
\F_{{\rm el}}^{*}(x)=-Q\grad\phi(x)\label{eq:Fel*}
\end{equation}
where $x$ is the molecule position and $Q$ the total charge on the
molecule. The drag force is more difficult, since it would actually
vanish for a point-sized molecule, but we still want a reasonable
estimate for a spherical molecule of finite radius $r>0$. We approximate
it by Stokes' law

\begin{equation}
\F_{{\rm drag}}^{*}(x)=6\pi\eta r\u(x).\label{eq:Fdrag*}
\end{equation}
Because of confinement inside the pore, we anticipate that using equation
\eqref{eq:Fdrag*} could lead to an underestimation of the drag force
\citep{Paine1975}. This is confirmed in Section \ref{sub:Explicit-vs.-implicit},
where the discrepancy between the two models $\F$ and $\F^{*}:=\F_{{\rm el}}^{*}+\F_{{\rm drag}}^{*}$
(finite-sized and point-sized molecule) is investigated. We shall
also elaborate on possible improvements to the point-size model.

\subsection{The Poisson-Boltzmann approximation}

Close to equilibrium, ion concentrations are well approximated by
the Boltzmann factors
\[
c^{\pm}=c_{0}e^{\mp\frac{q}{kT}\phi}
\]
where $c_{0}$ is the bulk concentration of both ion species. For
zero voltage bias, this is the exact solution to the Nernst-Planck
equations \eqref{eq:nernst-planck} without the convective flux term
$c^{\pm}\u$ (in this case, $\j^{\pm}=0$ and the boundary condition
$c^{\pm}=c_{0}$ is matched exactly if the potential is zero at the
boundary). But even for low to moderate biases and with convective
flux included, Boltzmann factors provide a useful estimate for the
ion distributions near charged walls. By plugging the Boltzmann distribution
into the Poisson equation \eqref{eq:poisson}, one obtains the Poisson-Boltzmann
(PB) equation

\[
-\nabla\cdot\left(\varepsilon\nabla\phi\right)+\chi_{F}2qc_{0}\sinh\frac{q\phi}{kT}=\rho_{0},
\]
where $\chi_{F}$ denotes the characteristic function of the fluid
domain. By using the linearization $\sinh(z)\approx z$, we arrive
at the \emph{linear }Poisson-Boltzmann equation\emph{
\begin{equation}
-\nabla\cdot\left(\varepsilon\nabla\phi\right)+\chi_{F}\frac{2q^{2}c_{0}}{kT}\phi=\rho_{0}.\label{eq:linearPB}
\end{equation}
}This gives us a simplified model for the electric potential. In Section
\ref{sub:Goal-oriented-adaptivity}, we construct an error indicator
solely based on solutions to the linear PB equation, which is orders
of magnitude cheaper than solutions of the full PNPS model. Thus,
we use equation \eqref{eq:linearPB} for mesh pre-refinement before
the actual PNPS simulation. The linearity also makes it fit better
into the framework of goal-oriented adaptivity.

In addition, the PB approximation can serve as initial guess for Newton
iteration of the PNP equations. This helps with convergence issues
caused by high surface charges, as shown in Section \ref{sub:linearization-comparison}.

\section{Numerical Method\label{sec:Numerical-Method}}

We use a finite element discretization for all the equations of the
PNPS system, which was implemented in Python on top of the open-source
finite element package Fenics \citep{Logg2012}. It allows us to easily
handle unstructured meshes on irregular geometries and a wide variety
of variational formulations and boundary conditions. %
The solver is released as a Python package, with all of our code made
available online.%
\footnote{\url{https://github.com/mitschabaude/nanopores}%
}

\subsection{Linearization of PNPS\label{sub:Linearization-of-PNPS}}

The PNPS system is nonlinear in the potential $\phi$, concentrations
$c^{\pm}$ and velocity $\u$ and therefore has to be solved by some
iterative linearization procedure. These procedures can be distinguished
by whether they use Newton's method, or fixed-point iteration, or
a combination. Along these lines we will describe three different
methods for the PNPS system. They will be compared numerically in
Section \ref{sub:linearization-comparison}.

\paragraph{Newton method}

The first, black box approach to linearization is to apply Newton's
method to the whole system. Let us write our nonlinear equation in
abstract variational form,
\[
F(U)=0,
\]
where $U=(\phi,c^{+},c^{-},\u,p)$ represents the finite element solution
and $F(U)$ is a linear functional on the discrete test space. Then,
Newton's method consists in successive linear solves of
\begin{equation}
DF(U)\delta U=-F(U),\label{eq:jacobian}
\end{equation}
followed by updates $U=U+\delta U$. As initial value, for instance,
$U=(\phi,c^{+},c^{-},\u,p)=(0,c_{0},c_{0},0,0)$ can be used. Written
out, equation \eqref{eq:jacobian} amounts to solving the system\begin{subequations}\label{eq:pnps-newton}
\begin{eqnarray}
-\Div(\varepsilon\grad\delta\phi)-F(\delta c^{+}-\delta c^{-}) & = & \rho_{0}+\Div(\varepsilon\grad\phi)+F(c^{+}-c^{-}),\\
\Div\left[-D^{+}\grad\delta c^{+}-\mu^{+}\delta c^{+}\grad\phi+\delta c^{+}\u-\mu^{+}c^{+}\grad\delta\phi+c^{+}\delta\u\right] & = & \Div\left[D^{+}\grad c^{+}+\mu^{+}c^{+}\grad\phi-c^{+}\u\right],\\
\Div\left[-D^{-}\grad\delta c^{-}+\mu^{-}\delta c^{-}\grad\phi+\delta c^{-}\u+\mu^{-}c^{-}\grad\delta\phi+c^{-}\delta\u\right] & = & \Div\left[D^{-}\grad c^{-}-\mu^{-}c^{-}\grad\phi-c^{-}\u\right],\\
-\eta\laplace\delta\u+\grad\delta p+F(\delta c^{+}-\delta c^{-})\grad\phi+F(c^{+}-c^{-})\grad\delta\phi & = & \eta\laplace\u-\grad p-F(c^{+}-c^{-})\grad\phi,\\
\Div\delta\u & = & -\Div\u,
\end{eqnarray}
\end{subequations}which is a linearized version of PNPS in the unknowns
$\delta U=(\delta\phi,\delta c^{+},\delta c^{-},\delta\u,\delta p)$.
Here we use $\mu^{\pm}:=\frac{qD^{\pm}}{kT}$ for readability. We
will refer to this algorithm as the \emph{(full) Newton method}:

\begin{algorithm}[H]
Initialize $U=(\phi,c^{+},c^{-},\u,p)$. Then:
\begin{enumerate}
\item Solve \eqref{eq:pnps-newton} for the Newton update $\delta U=(\delta\phi,\delta c^{+},\delta c^{-},\delta\u,\delta p)$.
\item Update $U=U+\delta U$.
\item Let the relative error be defined by
\[
\text{error}:=\frac{\n{\delta U}_{L^{2}(\Omega)}}{\n U_{L^{2}(\Omega)}}.
\]
Check convergence (with tolerance $\tau>0$):

\begin{enumerate}
\item If $\text{error}<\tau$, stop.
\item Else, go back to 1.
\end{enumerate}
\end{enumerate}
\caption{Newton method for PNPS\label{alg:newton}}
\end{algorithm}

\paragraph{Hybrid method}

The full Newton method yields a single large system for linear solving.
We suspected that computational effort could be saved by separating
the equations into two smaller subsets, namely the PNP and Stokes
systems, and solving them in an alternating fashion until convergence
is reached, i.e., by a fixed-point iteration. Thus, the potential
$\phi$ and concentrations $c^{\pm}$ will be obtained from the PNP
part of the system, with the initial velocity set to zero. Then, Stokes
will be solved for the velocity $\u$ and pressure $p$, with $\phi$,
$c^{+}$, $c^{-}$ as input; the velocity will now be plugged into
PNP, etc.

In this formulation, the Stokes part of the system is linear, but
we still need to address the remaining nonlinearity in the PNP part
(namely, the products $c^{\pm}\grad\phi$). We could again simply
apply Newton's method to the PNP system only, or use an inner fixed-point
loop. Both methods are well-established in the context of the PNP
equations without velocity term \citep{Lu2007,Lu2010}; however, the
fixed-point method is usually reported to require some form of under-relaxation,
resulting in several hundreds of iterations until convergence \citep{Lu2007}.

In practice, we initially experimented with a naive, unrelaxed fixed-point
iteration for PNP but found it to diverge in every case considered.
Newton iteration, applied to the PNP part only, was seen to be quite
robust, even with the constant initial guess $\phi=0$, $c^{\pm}=c_{0}$;
in situations where it does not converge either, resources such as
a better initial guess and under-relaxation (damped Newton) can still
be used. Regarding the fixed-point iteration between PNP and Stokes,
convergence was always observed; the coupling between these two systems,
which stems from the convective flux term $c^{\pm}\u$ in the Nernst-Planck
equation \eqref{eq:nernst-planck}, seems to be sufficiently weak.

Taking this into consideration, the second method we propose consists
of an outer fixed-point loop decoupling the system into a PNP and
a Stokes part, and an inner Newton iteration to solve the PNP part.
As a further refinement, it was found beneficial not to solve the
PNP equations to full convergence in every step, but only to perform
a single Newton iteration for every linear Stokes solve. We call this
version of the algorithm the \emph{hybrid method}:

\begin{algorithm}[H]
Initialize $U_{\text{PNP}}=(\phi,c^{+},c^{-})$ and $U_{\text{Stokes}}=(\u,p)$.
The Newton update equation for the PNP system is\begin{subequations}\label{eq:pnp-newton}
\begin{eqnarray}
-\Div(\varepsilon\grad\delta\phi)-F(\delta c^{+}-\delta c^{-}) & = & \rho_{0}+\Div(\varepsilon\grad\phi)+F(c^{+}-c^{-}),\\
\Div\left[-D^{+}\grad\delta c^{+}-\mu^{+}\delta c^{+}\grad\phi+\delta c^{+}\u-\mu^{+}c^{+}\grad\delta\phi\right] & = & \Div\left[D^{+}\grad c^{+}+\mu^{+}c^{+}\grad\phi-c^{+}\u\right],\\
\Div\left[-D^{-}\grad\delta c^{-}+\mu^{-}\delta c^{-}\grad\phi+\delta c^{-}\u+\mu^{-}c^{-}\grad\delta\phi\right] & = & \Div\left[D^{-}\grad c^{-}-\mu^{-}c^{-}\grad\phi-c^{-}\u\right].
\end{eqnarray}
\end{subequations}Now,
\begin{enumerate}
\item Solve \eqref{eq:pnp-newton} for the Newton update $\delta U_{\text{PNP}}=(\delta\phi,\delta c^{+},\delta c^{-})$.
\item Update $U_{\text{PNP}}=U_{\text{PNP}}+\delta U_{\text{PNP}}$.
\item Save $U_{\text{Stokes},0}=U_{\text{Stokes}}$ from the last iteration.\\Solve
the Stokes equation \eqref{eq:stokes} for $U_{\text{Stokes}}$.
\item Set $\delta U_{\text{Stokes}}=U_{\text{Stokes}}-U_{\text{Stokes},0}$.
Let the relative error be defined by
\[
\text{error}:=\frac{1}{2}\left(\frac{\n{\delta U_{\text{PNP}}}_{L^{2}(\Omega)}}{\n{U_{\text{PNP}}}_{L^{2}(\Omega)}}+\frac{\n{\delta U_{\text{Stokes}}}_{L^{2}(\Omega)}}{\n{U_{\text{Stokes}}}_{L^{2}(\Omega)}}\right).
\]
Check convergence (with tolerance $\tau>0$):

\begin{enumerate}
\item If $\text{error}<\tau$, stop.
\item Else, go back to 1.
\end{enumerate}
\end{enumerate}
\caption{Hybrid method for PNPS\label{alg:hybrid}}
\end{algorithm}

\paragraph{Fixed-point method}

Regardless of initial failure, we still wanted to see if a fast fixed-point
solution was possible. We took inspiration from a practice common
in the semiconductor community, which is to introduce the so-called
Slotboom variables \citep{Slotboom1973} 
\[
\tilde{c}^{\pm}=c^{\pm}e^{\pm\frac{q}{kT}\phi}.
\]
For motivation, note that in equilibrium -- when the Boltzmann distribution
applies -- $\tilde{c}^{\pm}$ reduce to constants. Plugging these
new variables into the classical Nernst-Planck equations (without
velocity) renders them in purely self-adjoint form, which is usually
emphasized \citep{Lu2007}. But, more importantly for us, it also
alters the form of the Poisson equation:

\[
-\nabla\cdot\left(\varepsilon\nabla\phi\right)=F(\tilde{c}^{+}e^{-\frac{q}{kT}\phi}-\tilde{c}^{-}e^{\frac{q}{kT}\phi}).
\]
Note that the potential $\phi$ now appears also on the right hand
side, in the charge distribution term. The Poisson equation becomes
nonlinear in the potential. If this would be solved in a fixed-point
loop, where $\tilde{c}^{\pm}$ are taken from the last iteration,
the new potential would not only give rise to a new charge distribution,
but would already reflect part of the impact this change has on itself.
This is just the correction that is needed to prevent the fixed-point
iteration from exploding. In semiconductor device modeling, this type
of iteration is known as \emph{Gummel's method }\citep{Gummel1964,Kerkhoven2006}.

With this new insight, we can actually change back to the old variables
$c^{\pm}$, by using $\tilde{c}^{\pm}=c^{\pm}e^{\pm\frac{q}{kT}\phi_{0}}$
with $\phi_{0}$ being the potential from the last iteration, to find
\[
-\nabla\cdot\left(\varepsilon\nabla\phi\right)=F(c^{+}e^{-\frac{q}{kT}(\phi-\phi_{0})}-c^{-}e^{\frac{q}{kT}(\phi-\phi_{0})}).
\]
In the limit of convergence of the fixed-point iteration we have $\phi=\phi_{0}$
and the exponential terms vanish. Since those are artificial correction
terms anyway, we might as well make our life easier and linearize
them as 
\[
e^{\pm\frac{q}{kT}(\phi-\phi_{0})}\approx1\pm\frac{q}{kT}(\phi-\phi_{0}).
\]
Reordering, we arrive at a corrected Poisson equation which is again
linear in the unknown $\phi$ and features the previous potential
$\phi_{0}$ on the right hand side:
\begin{equation}
-\nabla\cdot\left(\varepsilon\nabla\phi\right)+\frac{Fq}{kT}\phi(c^{+}+c^{-})=F(c^{+}-c^{-})+\frac{Fq}{kT}\phi_{0}(c^{+}+c^{-}).\label{eq:corrected-poisson}
\end{equation}
Alternating the corrected Poisson equation \eqref{eq:corrected-poisson}
with the unmodified Nernst-Planck and Stokes equations yields our
proposed \emph{fixed-point method}, summarized in Algorithm \ref{alg:fixed-point}.
To ensure convergence of the method, we recommend to start with two
pure PNP iterations and only afterwards include the Stokes equation
in the iteration.

\begin{algorithm}[H]
Initialize $\phi$, $c^{+}$, $c^{-}$ and $U_{\text{Stokes}}=(\u,p)$.
Set the iteration count to $i=0$. Then:
\begin{enumerate}
\item Save the solutions from the previous iteration: $\phi_{0}=\phi$,
$c_{0}^{+}=c^{+}$, etc.
\item Solve the corrected Poisson equation \eqref{eq:corrected-poisson}
for $\phi$.
\item Solve both the Nernst-Planck equations \eqref{eq:nernst-planck} for
$c^{+}$, $c^{-}$.
\item If $i<2$, set $i=i+1$ and go back to 1.
\item Solve the Stokes equation \eqref{eq:stokes} for $U_{\text{Stokes}}$.
\item Define the updates $\delta\phi=\phi-\phi_{0}$, etc. Let the relative
error be defined by\global\long\def\relerr#1{\frac{\n{\delta#1}_{L^{2}(\Omega)}}{\n{#1}_{L^{2}(\Omega)}}}
\[
\text{error}:=\frac{1}{4}\left(\relerr{\phi}+\relerr{c^{+}}+\relerr{c^{-}}+\relerr{U_{\text{Stokes}}}\right).
\]
Check convergence (with tolerance $\tau>0$):

\begin{enumerate}
\item If $\text{error}<\tau$, stop.
\item Else, go back to 1.
\end{enumerate}
\end{enumerate}
\caption{Fixed-point method for PNPS\label{alg:fixed-point}}
\end{algorithm}

\subsection{Details of the numerical solver\label{sub:Details-of-the-numerical}}

Variational formulations for the PNPS related systems were obtained
by integration by parts of equations \eqref{eq:nernst-planck}--\eqref{eq:stokes},
\eqref{eq:linearPB}, \eqref{eq:pnps-newton}--\eqref{eq:corrected-poisson}
respecting their conservative structure; this was done for both the
full 3D and 2D axisymmetric cases. The formulation of the Stokes system
in cylindrical coordinates can be found in \citep{Deparis2004}. Dirichlet
boundary conditions are enforced strongly, while Neumann conditions
are incorporated naturally via boundary terms on the right hand side.
For subequations of the PNP system, a standard $P^{1}$ finite element
discretization is employed, while for the Stokes system, the consistent
stabilized $P^{1}-P^{1}$ formulation by \citet{Hughes1987} is used
to ensure inf-sup stability; for some of the numerical experiments
below, we had to switch to a (more expensive) Taylor-Hood $P^{2}-P^{1}$
formulation for Stokes to enforce higher numerical accuracy.

In the 2D case, we always use a direct algebraic solver based on LU
decomposition. This quickly turned out to introduce a memory and speed
bottleneck in the 3D case, so we started experimenting with iterative
solvers based on Krylov subspace methods for both the PNP and Stokes
subsystems. The following choices are used subsequently for the 3D
solver with the hybrid method: The PNP system \eqref{eq:pnp-newton}
and PB equation \eqref{eq:linearPB}, which are elliptic, are solved
with BiCGstab \citep{Vorst2006} preconditioned by an incomplete LU
factorization, and the Stokes system with the transpose-free quasi-minimal
residual (TFQMR) method \citep{Freund2006} preconditioned by the
block preconditioner by \citet{Mardal2011}. For both solvers, the
Fenics interface to PETSc \citep{balay2014petsc} was employed, with
the ILU implementation taken from the Hypre package \citep{hypre}.

For the PNP system, the iterative approach drastically improves computational
speed in 3D even for moderately large meshes. For the Stokes system
however, a large number (hundreds to thousands) of iterations are
needed which renders the iterative approach quite slow, but at least
the memory barrier is removed. Further research will have to shed
light on constructing the right preconditioner for the Stokes system
in a nanopore context, and explain why an established method \citep{Mardal2011}
seems to fail here.

Meshes were generated with Gmsh \citep{Geuzaine2009} and the Python-Gmsh
interface written by Schlömer%
\footnote{\url{https://github.com/nschloe/pygmsh}%
}. Curved boundaries, as present in all our pore geometries, were approximated
by polygons. For mesh refinement, we modified the implementation in
Fenics so that refined meshes respect the curved geometry.

\subsection{Goal-oriented adaptivity\label{sub:Goal-oriented-adaptivity}}

Solutions to the PNPS system generally exhibit multiscale features
such as thin boundary layers near a charged surface. Overall simulation
accuracy depends on whether these layers are resolved by a sufficiently
fine mesh. However, since the main output of the simulation are scalar
quantities of interest like the force on a target molecule (Section
\ref{sub:Physical-Quantities}), the mesh has to be fine only in the
locations that directly influence this quantity. Especially in 3D,
the simulation time can quickly explode and get untractable if mesh
refinement is applied too broadly.

\begin{figure}
\begin{centering}
\includegraphics[width=0.4\textwidth]{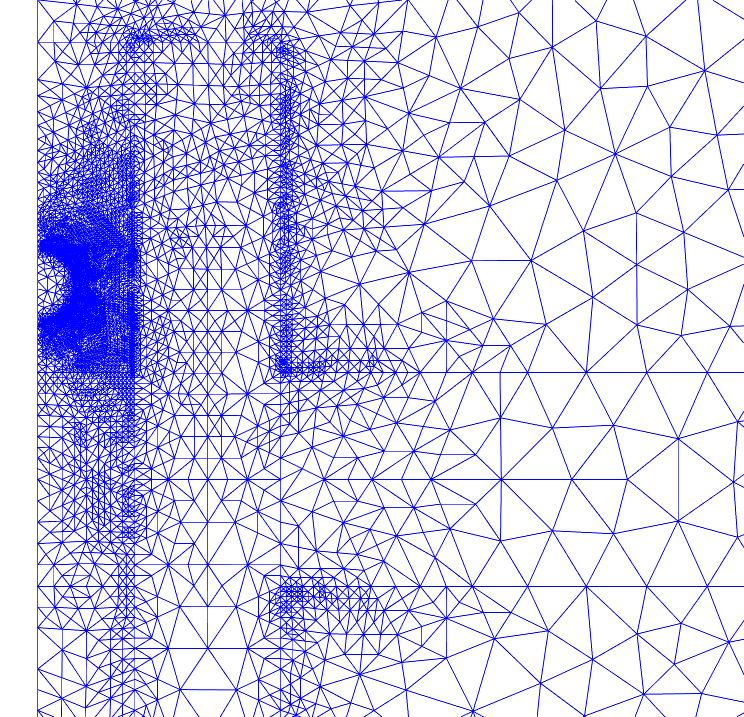}
\par\end{centering}

\caption{\label{fig:refined-example}Refined mesh produced by the goal-oriented
adaptive algorithm.}
\end{figure}

Figure \ref{fig:refined-example} shows an example mesh on an axisymmetric
pore geometry that fulfills these objectives: strong refinement at
the boundary layers with emphasis on the positions around the molecule,
but a coarse mesh outside the pore and far from the molecule. Clearly,
such an approach to mesh refinement would be cumbersome to implement
by hand and difficult to generalize. The method which accomplishes
this automatically is adaptive refinement based on goal-oriented error
estimation \citep{Becker2003,bangerth2013adaptive}, and will be described
in this section.

\paragraph{General framework}

To present the general ideas of goal-oriented adaptivity, suppose
we want to solve a linear PDE in variational form
\begin{equation}
a(u,v)=L(v).\label{eq:variational}
\end{equation}
Goal-oriented means that we are interested in a quantity $F(u)\in\R$,
where $F$ is a functional on the solution space and assumed to be
linear. The Galerkin discretization of \eqref{eq:variational} yields
a solution $u_{h}$ in a discrete subspace, and we want to refine
the mesh such that the error in the goal functional compared to the
true solution is small, i.e. decrease $|F(u-u_{h})|$. To analyze
this error we introduce the \emph{dual solution} $w$ which solves
\begin{equation}
a(v,w)=F(v).\label{eq:dual}
\end{equation}
Note that the arguments in the bilinear form have been swapped and
the right hand side is now the goal functional. Equations \eqref{eq:variational}
and \eqref{eq:dual} can be combined to give an error representation:
\[
F(u-u_{h})=a(u-u_{h},w)=L(w)-a(u_{h},w)=:R(w).
\]
The right-hand side is simply the residual of the Galerkin problem,
evaluated at the dual solution $w$, and can in principle be computed
once we have an approximation $w_{h}$ of $w$. Care must be taken
at this point, however, because choosing $w_{h}$ from the same discrete
space as $u_{h}$ yields $R(w_{h})=0$ due to Galerkin orthogonality,
which is hardly a useful error estimate. One possible solution is
to extrapolate $w_{h}$ patchwise to a higher-order function space
\citep{Rognes2012,bangerth2013adaptive}. In the case when $w_{h}$
and $u_{h}$ are $P^{1}$ and the extrapolation $Ew_{h}$ is $P^{2}$,
for example, one can expect that \citep{bangerth2013adaptive}
\[
F(u-u_{h})=R(w)=O(h^{2})\quad\text{while}\quad R(Ew_{h})=R(w)+O(h^{3}).
\]
This shows that $F(u-u_{h})\approx R(Ew_{h})$ produces an accurate
estimate of the true error. A local error indicator is obtained by
expanding the residual into a sum of local contributions,
\[
R(w)=\sum_{T}R_{T}(w)+R_{\partial T}(w).
\]
The local terms can be found by integration by parts on every mesh
element $T$. Their absolute values 
\begin{equation}
\eta_{T}:=\left|R_{T}(\tilde{w})+R_{\partial T}(\tilde{w})\right|\label{eq:indicators}
\end{equation}
are the element-wise error indicators. Here we will use $\tilde{w}=Ew_{h}-w_{h}$
to make the residuals as small and the upper bound $|F(u-u_{h})|\le\sum\eta_{T}$
as tight as possible. The error indicators $\eta_{T}$ \eqref{eq:indicators}
are used to decide which mesh elements will be refined, as outlined
in the following classical adaptive algorithm.

\begin{algorithm}[H]
Starting with an initial coarse mesh, do:
\begin{enumerate}
\item Solve the primal \eqref{eq:variational} and dual \eqref{eq:dual}
problems for $u_{h}$, $w_{h}$.
\item Extrapolate $w_{h}$ to a higher-order approximation $Ew_{h}$.
\item Compute the error indicators \eqref{eq:indicators}.
\item Produce a refined mesh by bisecting the elements with the highest
indicators.
\item Based on some criterion, either stop or go back to 1.
\end{enumerate}
\caption{Adaptive algorithm\label{alg:adaptive}}

\end{algorithm}
The patchwise extrapolation in step 2 is done using the Fenics implementation
detailed by \citet{Rognes2012}. The number of elements to be refined
in step 4 is determined by Dörfler marking \citep{Dorfler1996}.
As for the termination criterion in step 5, we will usually stop after
a critical amount of elements is reached corresponding to a maximal
work load.

\paragraph{Faster error estimation without extrapolation}

In our implementation of the adaptive Algorithm \ref{alg:adaptive},
the most computationally expensive step is extrapolation of the dual
solution to a higher-order function space. We expect this observation
to hold independently of the precise extrapolation method, due to
the blow-up of degrees of freedom when switching from a $P^{1}$ to
a $P^{2}$ representation. As discussed, the extrapolation is done
to avoid Galerkin orthogonality, improving the useless global estimate
$F(u-u_{h})\approx R(w_{h})=0$ to an estimate of (sometimes provable)
high accuracy $F(u-u_{h})\approx R(Ew_{h})$. However, that this also
leads to local error indicators of high quality remains largely a
heuristic, according to current wisdom about adaptive methods \citet{feischl2016abstract}.

In any case, we will apply error estimation only in an approximate
way to a simplified model (explained below). One consequence is that
the \emph{global} error estimate we would get from the adaptive algorithm
has no relevance to our actual quantities of interest. Once we dispense
with a global error estimate and only want to compute local indicators,
there is no longer a clear theoretical motivation to use the extrapolated
dual solution $Ew_{h}$ rather than $w_{h}$. Consequently, we propose
to exchange the indicators \eqref{eq:indicators} for
\begin{equation}
\eta_{T}^{{\rm cheap}}:=\left|R_{T}(w_{h})+R_{\partial T}(w_{h})\right|.\label{eq:indicators-cheap}
\end{equation}
Thus, we simply skip the extrapolation step and calculate our error
indicators directly from the dual Galerkin solution. This leads to
the following alternative adaptive Algorithm \ref{alg:adaptive-cheap},
which is both faster and simpler to implement than Algorithm \ref{alg:adaptive}.
\begin{algorithm}[H]
Starting with an initial coarse mesh, do:
\begin{enumerate}
\item Solve the primal \eqref{eq:variational} and dual \eqref{eq:dual}
problems for $u_{h}$, $w_{h}$.
\item Compute the cheap error indicators \eqref{eq:indicators-cheap}.
\item Produce a refined mesh by bisecting the elements with the highest
indicators.
\item Based on some criterion, either stop or go back to 1.
\end{enumerate}
\caption{Cheap adaptive algorithm without extrapolation\label{alg:adaptive-cheap}}
\end{algorithm}
Interestingly, we are not aware of any publication that makes use
of this simple idea, although it has also been proposed in the unpublished
work \citep{Jansson2013}. The performance with respect to quantities
of interest for our particular problem remains to be examined in Section
\ref{sub:Assessment-of-goal-adaptive}.

\paragraph{Adaptivity for the Poisson-Boltzmann equation}

Because two systems of equations have to be solved in every iteration
of Algorithms~\ref{alg:adaptive} and \ref{alg:adaptive-cheap},
running it for the full PNPS model would be costly. Besides, as a
nonlinear system it does not fit neatly into the framework sketched
above (but see \citep{bangerth2013adaptive} for goal-oriented adaptivity
in the nonlinear case). However, we expect a good deal of the error
information to be already contained in a simplified physical model,
namely, the linear PB equation. In our simulations, we will therefore
proceed as follows:\begin{enumerate}[Step 1.]
\item Refine the mesh several times by goal-oriented adaptivity for the linear PB equation \eqref{eq:linearPB}.
\item Solve the PNPS equations \eqref{eq:nernst-planck}--\eqref{eq:stokes} once on the final mesh.
\end{enumerate}When we apply the general adaptivity framework to the linear PB equation,
the bilinear form becomes
\[
a(\phi,\psi)=\int_{\Omega}\varepsilon\grad\phi\cdot\grad\psi+\int_{\Omega}\kappa\phi\psi
\]
with $\kappa:=\chi_{F}\frac{2q^{2}c_{0}}{k_{B}T}$ and $\chi_{F}$
the characteristic function of the fluid domain. The right hand side
stems from a composition of volume and surface charges
\[
L(\psi)=\int_{\Omega}\rho_{0}\psi+\int_{\Gamma}\rho\psi
\]
with $\Gamma$ denoting all charged interfaces, e.g., pore walls.
Both trial and test functions are restricted to vanish on the Dirichlet
part of $\partial\Omega$. That is, even if the full PNPS model is
specified with non-zero voltage bias, we always solve the auxiliary
PB equation with zero voltage bias, since applying a non-zero bias
would violate the assumptions underlying the PB model, leading to
an unphysical, exponential behaviour of the potential at the Dirichlet
boundary and an overestimation of error near this boundary.

The important remaining question is how to choose the goal functional.
Ideally, our quantity of interest would be the force $\F$ on the
target molecule \eqref{eq:Ftotal}, but computing it requires the
full PNPS solution. Still, we may find a functional of the linear
PB potential that roughly preserves the properties of the PNPS force,
such as higher relevance of the region near the molecule. We aim to
achieve this by defining 
\begin{equation}
F(\phi):=-\int_{M}\rho_{0}\partial_{z}\phi\label{eq:goal-PB}
\end{equation}
as our goal functional; $M$ is the molecule region and $\rho_{0}$
its charge density. This is a direct analogue of the electrical force~\eqref{eq:Fel}
from the PNPS model; the electric field $-\grad\phi$ is restricted
to the $z$-direction because we need a scalar-valued functional.
The hydrodynamic contribution to the force is neglected for lack of
a reasonable analogue in the PB model. 

It has to be emphasized that \eqref{eq:goal-PB} shall \emph{not}
provide a quantitative estimate of the actual force $\F$ or its electrical
part $\F_{\text{el}}$. The only purpose of the goal functional is
to guide the adaptive mesh refinement process, and we must hope that
the parts of the domain where the discretization error $F(\phi-\phi_{h})$
is high are the same parts where also the error in the actual\emph{
}quantity of interest $\F$ would be high. That this is indeed the
case is verified experimentally in Section~\ref{sub:Assessment-of-goal-adaptive}.

With these specializations of the general goal-adaptive framework,
let $w$ denote the dual solution as before. The local error contributions
can be obtained from the residual $R(w)$ via element-wise integration
by parts by calculating
\begin{align*}
R(w)=-a(\phi_{h},w)+L(w)= & \sum_{T}-\int_{T}\left(\varepsilon\grad\phi\cdot\grad w+\kappa\phi w\right)+\int_{T}\rho_{0}w+\frac{1}{2}\int_{\partial T}\rho w\\
= & \sum_{T}\int_{T}\left(\varepsilon\grad^{2}\phi-\kappa\phi+\rho_{0}\right)w+\frac{1}{2}\int_{\partial T}\left(\rho-[\varepsilon\partial_{n}\phi]\right)w\\
=: & \sum_{T}(r_{T},w)_{T}+(r_{\partial T},w)_{\partial T}.
\end{align*}
The strong cell and facet residuals are given by
\begin{eqnarray*}
r_{T} & = & \varepsilon\grad^{2}\phi-\kappa\phi+\rho_{0},\\
r_{\partial T} & = & \frac{1}{2}\left(\rho-[\varepsilon\partial_{n}\phi]\right),
\end{eqnarray*}
where $[\varepsilon\partial_{n}\phi]$ denotes the jump over a facet
in normal direction. Here, $\rho$ is understood to be zero on uncharged
facets, and on boundary facets $[\varepsilon\partial_{n}\phi]=\varepsilon\partial_{n}\phi$.
The error indicators $\eta_{T}$, defined abstractly in \eqref{eq:indicators}
and \eqref{eq:indicators-cheap}, can now be explicitly written as
\[
\eta_{T}=\left|(r_{T},\tilde{w})_{T}+(r_{\partial T},\tilde{w})_{\partial T}\right|
\]
with either $\tilde{w}=(Ew_{h}-w_{h})$ or $\tilde{w}=w_{h}$. Goal-oriented
adaptivity is implemented according to Algorithms~\ref{alg:adaptive}
and \ref{alg:adaptive-cheap}. Once the adaptive loop has produced
a mesh sufficiently resolving the regions of interest, the final Poisson-Boltzmann
solution is used as an initial guess for the iteration solving the
nonlinear PNPS equations.

\section{Results and Discussion\label{sec:Results-and-Discussion}}

In this section, we want to assess our numerical approach in detail
and gain further insights into the modeling and simulation. To embed
numerical experiments in a real-world application setting, we introduce
a 3D model of the nanopore sensor recently published in \citep{Burns2016}.

For a basic validation of the correctness of our implementation, we
also carried out experiments on an idealized test problem where a
semi-analytical solution is possible. They can be found in the Appendix
A.1.

\subsection{DNA-based nanopore sensor\label{sub:Model-of-DNA-based}}

\begin{figure}
\begin{centering}
\includegraphics[width=0.55\textwidth]{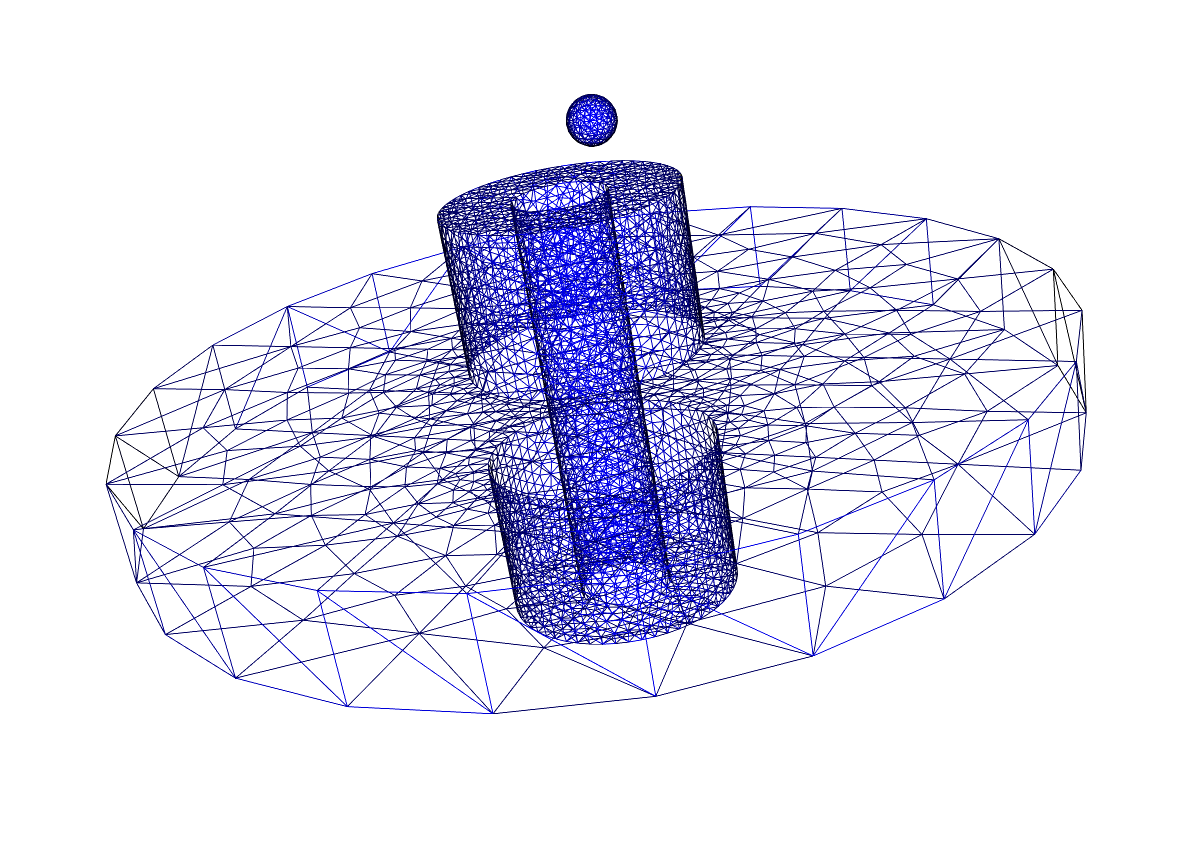}\qquad{}\includegraphics[width=0.3\textwidth]{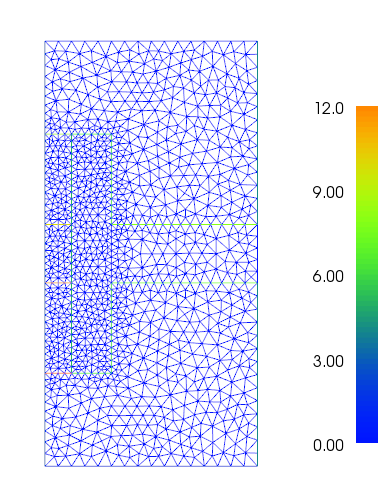}
\par\end{centering}

\caption{\label{fig:simple-pore}DNA nanopore model. Shown are the 3D mesh
of the pore, membrane and molecule, and the 2D mesh of a half-plane
on which the problem is solved in the axisymmetric formulation. The
geometry is inspired by \citep{Burns2016}.}
\end{figure}

We describe the DNA-based nanopore sensor model inspired by \citep{Burns2016}.
The pore consists of six vertically aligned $9$ nm short DNA strands
arranged in a circle, forming a hole of width approximately $5$ nm.
As shown in Fig.~\ref{fig:simple-pore}, we modeled the DNA pore
wall and hole by two concentric cylinders of radii $2.5$ and $1\text{ nm}$,
respectively (thus assuming the DNA thickness to be $1.5$ nm on average).
The pore resides in a lipid bilayer membrane of thickness $2.2$ nm,
and the whole system is placed in the center of a cylindrical electrolyte
reservoir of height $20$ nm and radius $10$ nm. We use relative
electrical permittivities $\varepsilon_{r}=80.2$ for water, $\varepsilon_{r}=2$
for the lipid bilayer and $\varepsilon_{r}=12$ for DNA. In some of
the following numerical experiments, we also include a spherical target
molecule, as depicted in Fig.~\ref{fig:simple-pore}, which has $\varepsilon_{r}=12$.
Without an off-centered molecule, the domain is axisymmetric and simulations
can optionally be done with the 2D solver.

Concerning boundary conditions, the DNA surface (inner and outer pore
wall, except the outer part which touches the membrane) is equipped
with a uniform negative surface charge density $\rho_{\text{DNA}}$,
while the membrane was assumed uncharged; a potential bias $\Delta V$
is applied on the bottom of the domain; and both ion concentrations
are fixed to a bulk value $c_{0}$ on the top and bottom. If not stated
otherwise, the numerical values used are $\rho_{\text{DNA}}=-0.25$$q/\text{nm}^{2}$,
$\Delta V=-100$ mV and $c_{0}=300$ $\text{mol}/\text{m}^{3}$. The
molecule is equipped with a fixed total charge of $-q$, which is
spread uniformly across the volume occupied by the discretized molecule.

\subsection{Comparison of linearization schemes\label{sub:linearization-comparison}}

In Section \ref{sub:Linearization-of-PNPS}, we proposed three approaches
for linearization of the PNPS system, which we now compare regarding
convergence speed and robustness. We consider the axisymmetric (2D)
formulation without target molecule as pictured in Fig.~\ref{fig:simple-pore}
(right), with a fixed quasi-uniform mesh size with $h\approx0.1$
nm.

\begin{figure}
\begin{centering}
\begin{minipage}[t]{0.45\textwidth}%
\subfigure[Linearization error vs. iteration count]{\resizebox{1.\textwidth}{!}{\includegraphics[width=1\paperwidth]{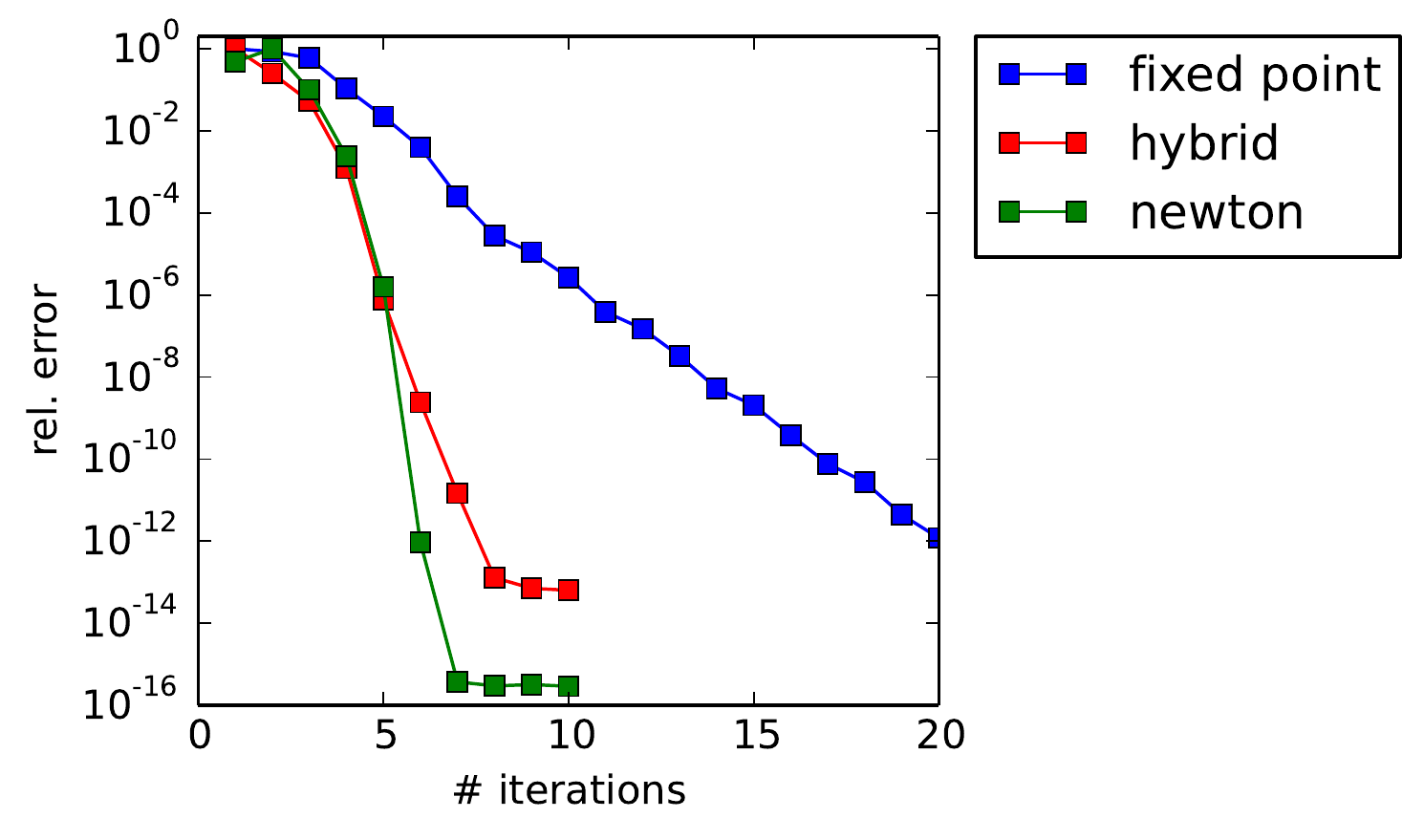}\label{fig:hybrid-vs-newton-iter}}}%
\end{minipage}\hspace*{\fill}%
\begin{minipage}[t]{0.45\textwidth}%
\subfigure[Linearization error vs. CPU time]{\resizebox{1.\textwidth}{!}{\includegraphics[width=1\paperwidth]{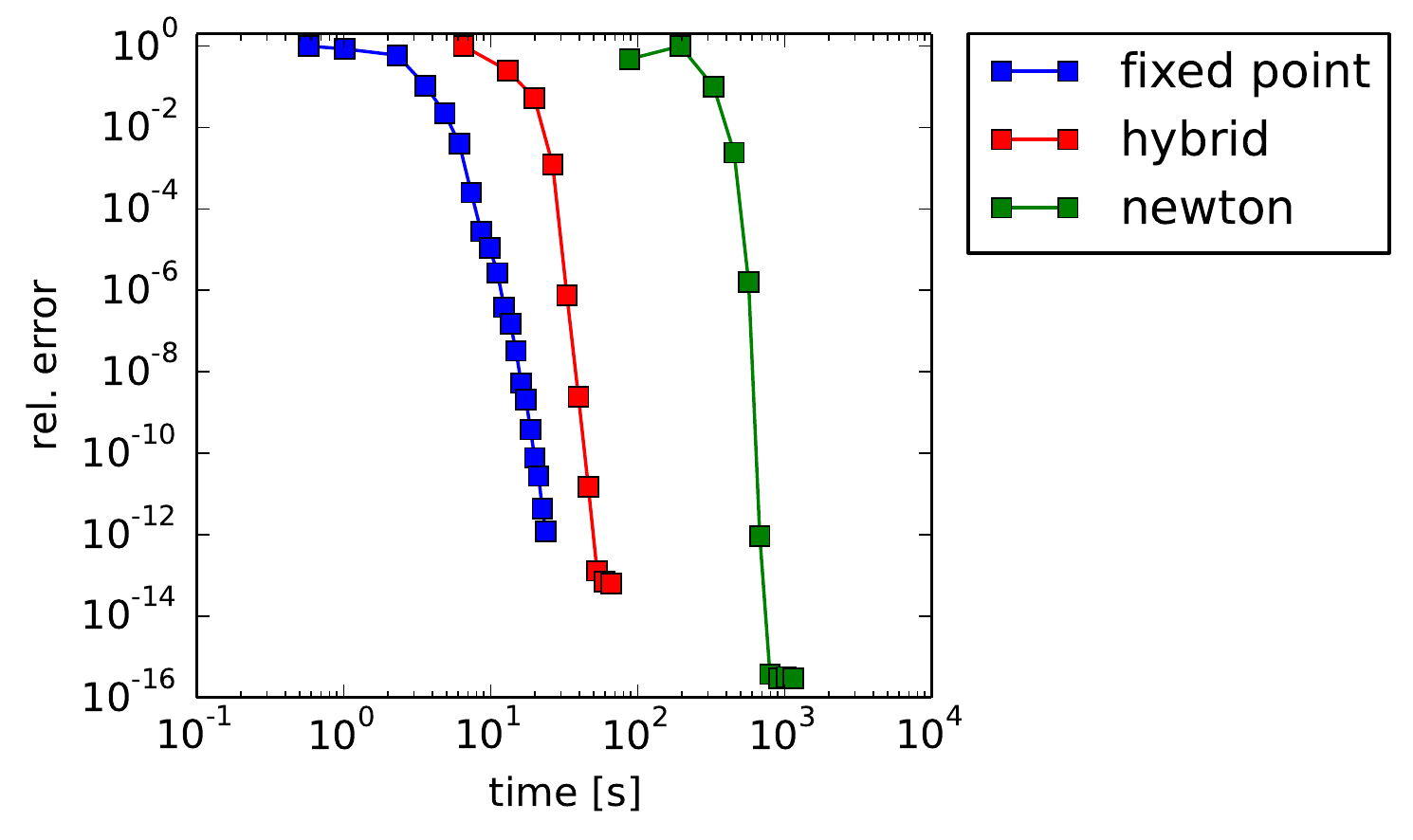}\label{fig:hybrid-vs-newton-cpu}}}%
\end{minipage}
\par\end{centering}

\centering{}\caption{\label{fig:hybrid-vs-newton}Comparison of the linearization error
between the Newton, hybrid and fixed-point methods. The error is
computed as defined in Algorithms \ref{alg:newton}, \ref{alg:hybrid}
and \ref{alg:fixed-point}, respectively. }
\end{figure}

\paragraph{Comparison of convergence speed}

Figure \ref{fig:hybrid-vs-newton-iter} compares convergence speed
of the Newton, hybrid and fixed-point methods in terms of the number
of iterations. The voltage bias was lowered to $-50$ mV in this example
to ensure convergence of all three schemes. The two Newton-based methods
exhibit fast, superlinear convergence at first; the hybrid method
deviates from this only in the last few steps and settles for the
expected linear asymptotic rate. For the fixed-point method, a slower
linear convergence rate can be observed. This indicates that the nonlinearity
in the PNP system is far stronger than the coupling between PNP and
Stokes, and the iteration between the latter two converges very fast. 

In terms of computational time, the picture is turned upside down,
because a single iteration of the fixed point scheme is much faster
than for the other schemes. All in all, as shown in Figure \ref{fig:hybrid-vs-newton-cpu},
the fixed point method clearly wins in terms of convergence speed.
If we compare the times when an accuracy of $4$ digits is first reached,
fixed-point (with $8.6$ seconds) is $3.8\times$ faster than hybrid
($33$ seconds) and $67\times$ faster than Newton (over $9$ minutes)
in this particular example.%
{} On smaller meshes than shown in the figure, all three methods are
closer, meaning the segregated approaches scale better with mesh size.

To put these results into the right perspective, let us note that
in Figure \ref{fig:hybrid-vs-newton-cpu}, computational times are
dominated by the algebraic linear solver, and that we used a direct
method for solving -- which has $O(N^{3/2})$ complexity in the number
of variables \citep{george1981computer} and thus gives an inherent
advantage to decoupled systems, as $2(N/2)^{3/2}<N^{3/2}$. This could
be considered an unfair comparison and raises the question whether
the advantage remains if an iterative solver of optimal complexity
$O(N)$ is used. However, choosing a good preconditioner for the full
Newton system seems a more difficult task than for the segregated
solvers. For the hybrid and fixed-point methods, the PNP and Stokes
subsystems can be treated separately, which makes them more convenient
when an iterative solver is used.

In the case that an LU solver is used, like here, there is an additional
tweak for the fixed-point and hybrid method that we have not mentioned:
Since the bilinear form of the Stokes equations is not changing, the
system matrix has to be assembled and factorized only once. This makes
the dominance of the fixed point method even more pronounced, but
for comparison's sake it was not used in Figure \ref{fig:hybrid-vs-newton-cpu}.

\begin{figure}
\begin{centering}
\begin{minipage}[t]{0.45\textwidth}%
\includegraphics[height=5.8cm]{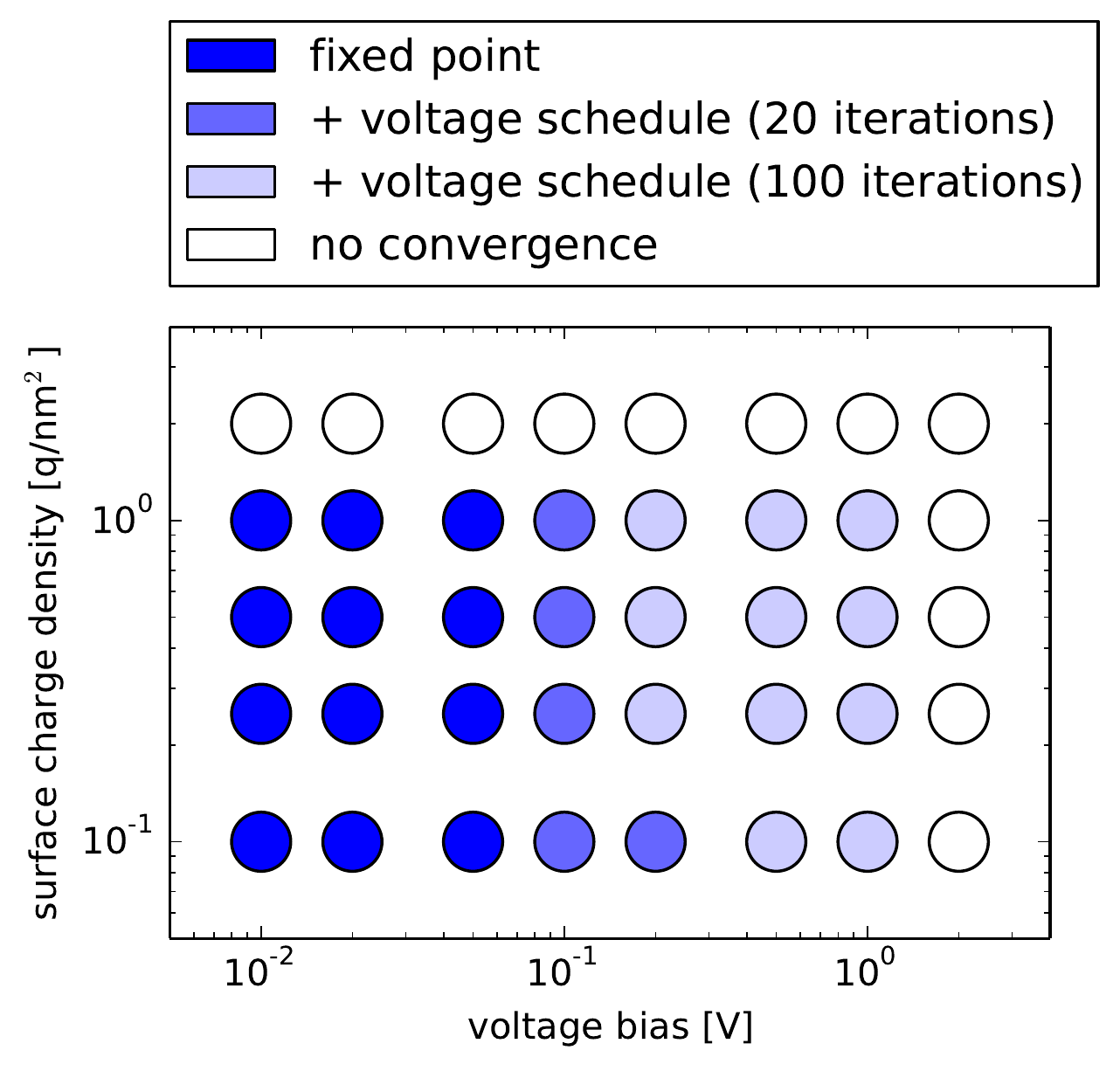}\label{fig:fixed-point-robustness}%
\end{minipage}\hfill{}%
\begin{minipage}[t]{0.45\textwidth}%
\includegraphics[height=5.8cm]{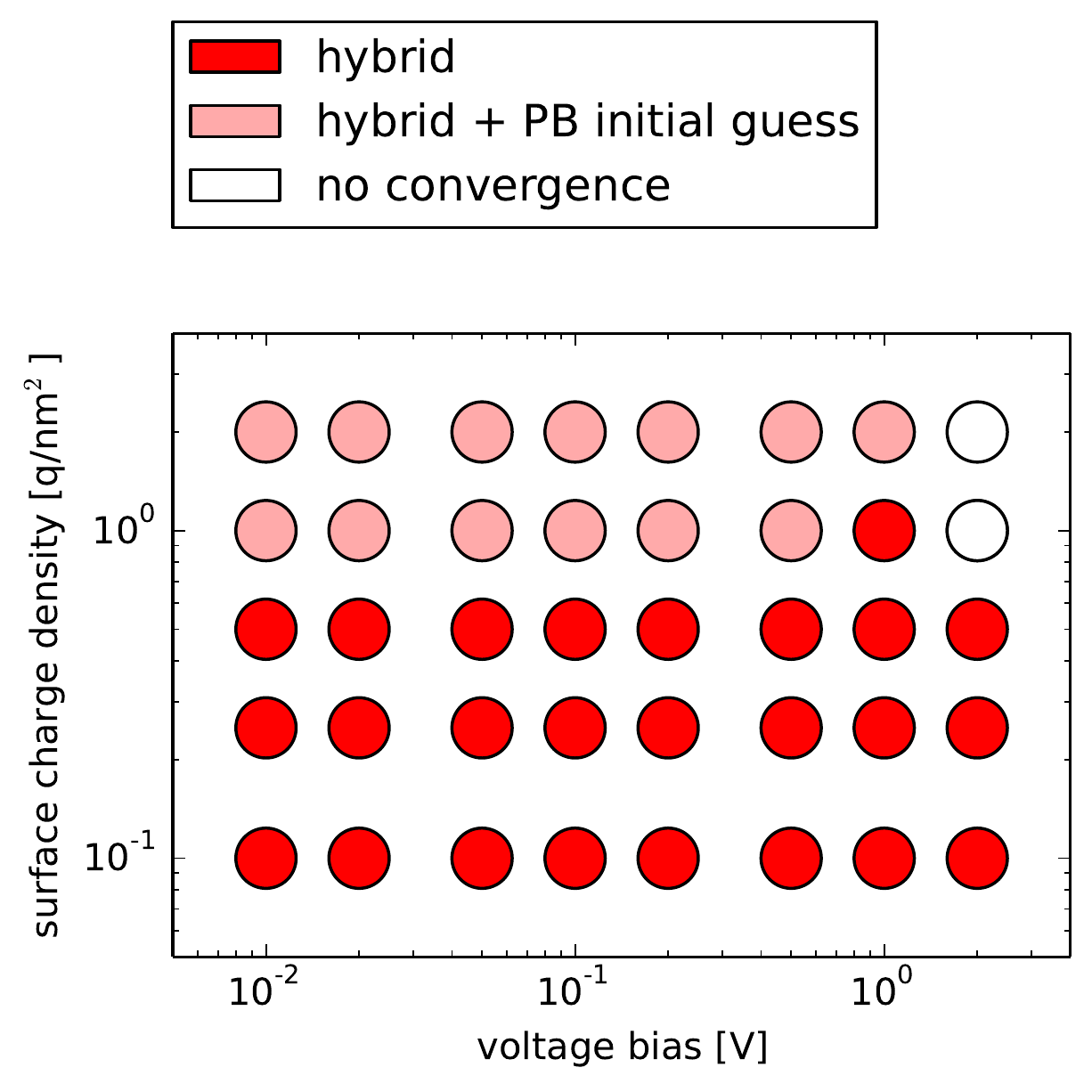}\label{fig:hybrid-point-robustness}%
\end{minipage}
\par\end{centering}

\centering{}\caption{Robustness of convergence of hybrid and fixed-point schemes with respect
to voltage bias/surface charge. The colored circles represent values
of voltage and surface charge where the respective iteration converges.
Convergence is defined as the linearization error (see Figure \ref{fig:hybrid-vs-newton})
falling below $\text{tol}=10^{-3}$; in case of non-convergence the
solution either exploded or, for of the fixed-point methods with voltage
schedule, an allowed number of iterations was exceeded. Lighter colors
are meant to subsume darker colors; i.e. the \emph{hybrid method +
PB initial guess} converges for same values as the \emph{hybrid method}
(red) plus additional values which are colored in light red. A white
circle indicates that none of the respective methods converged. \label{fig:robustness}}
\end{figure}

\paragraph{Comparison of convergence robustness}

Next, we compare robustness of convergence with respect to variations
of the model parameters. In the speed comparison, parameters where
chosen deliberately so that all three linearization methods converged.
But, as we will show, either of the methods diverges if some input
parameters are made large enough. We want to explore and be aware
of these convergence boundaries, as far as they are relevant to realistic
models.

The two parameters we identified as critical for convergence are surface
charge and voltage bias. Interestingly, the hybrid method is mainly
sensitive to surface charge but not to voltage bias, while for the
fixed-point method the opposite is true. In Figure \ref{fig:robustness},
we illustrate the region of convergence for these two methods in the
2D plane of DNA surface charge and voltage bias parameters, with the
(negative) surface charge density ranging up to $2q/\text{nm}^{2}$
and the (negative) voltage%
\footnote{For DNA-based nanopore models, we expect $\pm0.2$ V to be sufficient,
since for larger voltages the pore is not stable anyway \citep{Burns2013}.
The surface charge of DNA is about $-q/\text{nm}^{2}$, which is
already high compared to other materials.%
} up to $2$ V. We disregard the full Newton method at this point because
of its unfavourable speed performance, and because we observed its
robustness properties to be similar to the hybrid method.

The parameter values for which convergence is achieved are colored
in Figure \ref{fig:robustness} in bright blue and red for the fixed-point
and hybrid method, respectively. As can be seen, the fixed-point method
has problems even for a moderate voltage bias of $-0.1$ V, while
the hybrid method converges for all voltage biases. On the other hand,
the hybrid method has trouble with a (realistic) DNA charge density
of $-q/\text{nm}^{2}$.

In light of these observations, we propose a simple modification to
either of the two methods to migitate convergence issues. For the
fixed-point method, the idea is to slowly ramp up the voltage during
the iteration, beginning with zero and increasing the voltage bias
by a fixed small amount per step until the desired value is reached.
Until the iteration has arrived at the full voltage, only the PNP
equations are solved. A step-wise increase of $0.025$ V has been
found to work well; higher values can cause the iteration to become
unstable and diverge. With this modification, the fixed-point method
converges for every considered voltage bias, but the number of iterations
can become unsatisfyingly high if a large number of steps is needed
to divide the voltage by $0.025$ V. Therefore, in Figure \ref{fig:robustness},
we only count the iteration as converged if the desired tolerance
of $10^{-3}$ was reached during a specified number of iterations,
$20$ or $100$ respectively. The modified fixed-point algorithm is
denoted as \emph{fixed point + voltage schedule }in the figure.

For the hybrid method, which has problems with high surface charge,
we modify it by first solving the nonlinear Poisson-Boltzmann equation
and using that as an initial guess. This makes sense because the stationary
exponential distribution of ions inside the Debye layer is well captured
by the PB model, and is exactly what causes the Newton method for
PNP to diverge. The PB equation is solved with zero voltage bias;
the resulting potential $\phi$ and the ion concentrations $c^{\pm}=c_{0}e^{\mp\frac{q}{kT}\phi}$
provide the initial guess; the Stokes variables $\u$ and $p$ are
initialized to zero as before. The additional nonlinear PB solve adds
only small computation overhead to the much larger PNPS system. As
we can see in Figure~\ref{fig:robustness}, this modification greatly
enhances the robustness of the hybrid method: the \emph{hybrid method
+ PB initial guess }converges for almost all parameter values considered.
It is more robust than any variant of the fixed-point method and is
arguably the method of choice if a large range of voltage biases is
needed.

Both linearization methods could be made even more robust by using
some form of damping or under-relaxation, which we avoided here because
damping inherently increases the number of iterations. An adaptive
scheme, for example in case of the Newton iteration that is part of
the hybrid method, to using damping only when required and the solution
is in danger of diverging would definitely be useful to further increase
robustness.

In the remaining numerical experiments, the hybrid method is always
used.

\subsection{Assessment of goal-adaptive refinement strategy\label{sub:Assessment-of-goal-adaptive}}

Next, we investigate the goal-oriented adaptive scheme from Section
\ref{sub:Goal-oriented-adaptivity}. We use the same geometry as in
the last subsection, but with a spherical target molecule of radius
$0.5$ nm inserted at $z=2$~nm, inside the pore, slightly above
the center. The initial mesh is coarse and quasi-uniform, similar
to the ones depicted in Fig.~\ref{fig:simple-pore}.

We are interested if the force on the target molecule from the PNPS
system is computed with increasing accuracy on the meshes produced
by our adaptive algorithms. \emph{A priori}, this is not clear because
the estimator is tailored towards the auxiliary linear PB model with
an unphysical pseudo-force as goal functional; it does not necessary
yield good results when applied to the full PNPS model. For computation
of the error, we created a very fine reference solution with the axisymmetric
2D solver; the actual experiments were carried out with both the 2D
and 3D solvers.

Since our geometry has curved parts (spherical molecule and, in 3D,
cylindrical pore) whose polygonial discretization influences the numerical
values of goal functionals, we adapt the mesh in such a way that new
vertices created on boundary facets lie on the actual curved boundaries
rather than on the initial discrete polygons. In other words, after
mesh bisection the geometry of the problem is slightly altered in
every step of the adaptive loop. This is necessary to ensure convergence
of a 3D discretization to an axisymmetric 2D reference solution (and,
finally, to the actual continuum solution). On the other hand, the
geometric error cannot be captured by our error estimator, so performance
of the adaptive algorithm may be worse than predicted by theory. Another
remark concerning the adaptation process is that we always choose
volume and surface charge densities so that the \emph{total charge}
on molecule and pore wall are the same as they would be for the exact
curved geometry (i.e., we change the densities in every step), because
this makes more sense from a physical point of view.

\begin{figure}
\begin{centering}
\begin{minipage}[t]{0.45\textwidth}%
\subfigure[Convergence of force: 2D]{\resizebox{1.\textwidth}{!}{\includegraphics[width=1\paperwidth]{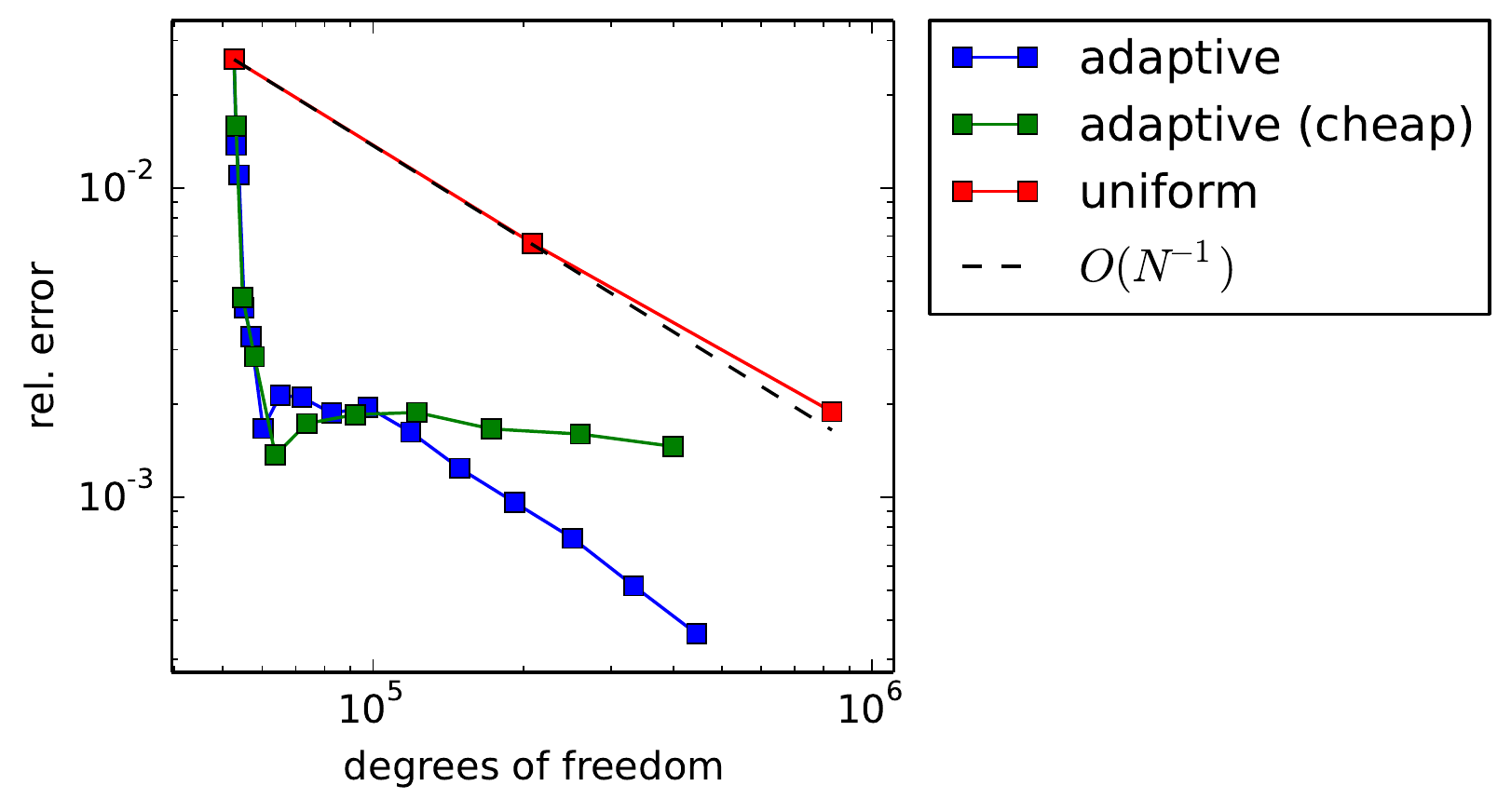}\label{fig:adap2D}}}%
\end{minipage}\hfill{}%
\begin{minipage}[t]{0.45\textwidth}%
\subfigure[Convergence of force: 3D]{\resizebox{1.\textwidth}{!}{\includegraphics[width=1\paperwidth]{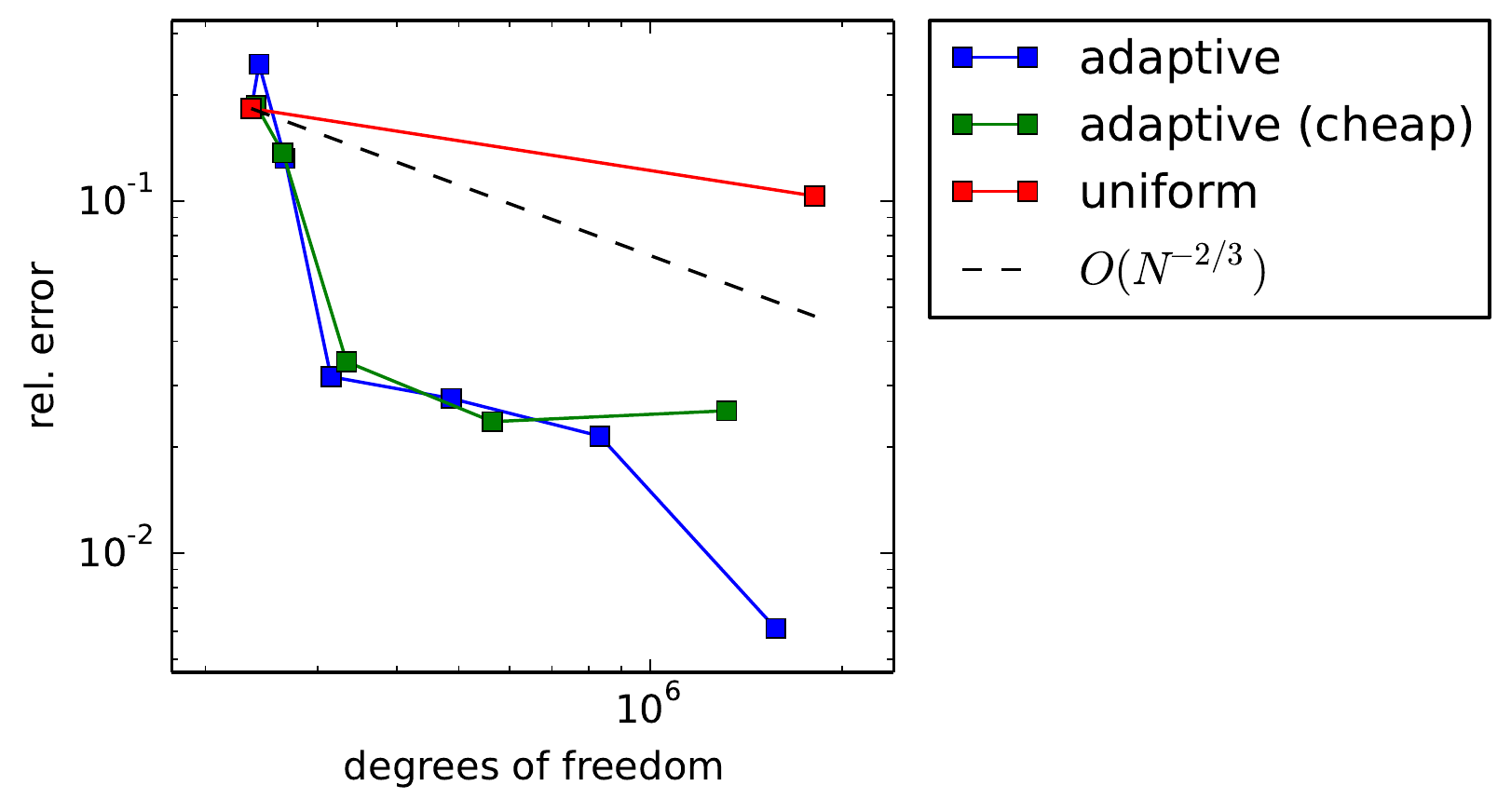}\label{fig:adap3D}}}%
\end{minipage}
\par\end{centering}

\captionsetup{singlelinecheck=off}
\caption[.]{Performance of different refinement strategies with respect to the
force \eqref{eq:Ftotal} in axisymmetric 2D (left) and 3D (right).
Starting from the same initial mesh, we compare the classical adaptive
Algorithm \ref{alg:adaptive}, the cheap adaptive Algorithm \ref{alg:adaptive-cheap}
and uniform refinement of all elements. Shown is the error relative
to a fine 2D reference solution, defined as
\[
{\rm error}:=\frac{|\F_{{\rm el},z}-\F_{{\rm el},z}^{{\rm ref}}|}{|\F_{{\rm el},z}^{{\rm ref}}|}+\frac{|\F_{{\rm drag},z}-\F_{{\rm drag},z}^{{\rm ref}}|}{|\F_{{\rm drag},z}^{{\rm ref}}|}.
\]
We compare only the $z$-components of the forces since the radial
components are zero by design in the axisymmetric formulation (the
molecule sits on the central axis). The reason we sum absolute values
of electrical and drag error components is to avoid a ``lucky cancellation''
that was observed in several data points and would make the errors
shown in this figure unreliable.\label{fig:adap}}
\end{figure}

For the PNPS force in 3D, it turned out that we could not exhibit
satisfying convergence when using the stabilized $P^{1}-P^{1}$ formulation
of the Stokes system (results not shown). This was remedied by switching
to a Taylor-Hood ($P^{2}-P^{1}$) formulation; the additional error
introduced by the stabilization seems to dominate the pure discretization
error on coarse meshes. The results shown in Figure \ref{fig:adap}
were obtained with the Taylor-Hood formulation.

In Figure \ref{fig:adap}, we compare the different refinement strategies
proposed in Section \ref{sub:Goal-oriented-adaptivity}, the classical
goal-adaptive algorithm with extrapolation and the proposed cheaper
variant without extrapolation, as well as uniform refinement as a
baseline. Both in 2D (left) and 3D (right), the adaptive schemes are
clearly superior to uniform refinement, especially in the first couple
of steps when the error gets quickly reduced by about one order of
magnitude. After that, however, the cheap estimator degrades in performance,
while the classical estimator seems to enter an asymptotic phase where
a well-defined convergence rate of is assumed, which can be heuristically
derived from theory using \emph{a priori} estimates%
\footnote{\label{note1}To derive the theoretical rates, consider a goal functional
$F(u)$, a bilinear form $a(\cdot,\cdot)$ and a discrete solution
$u_{h}$. Let $w$ and $w_{h}$ be the continuous resp. discrete dual
solutions as in Section~\ref{sub:Goal-oriented-adaptivity}. The
error can be estimated as
\[
F(u-u_{h})=a(u-u_{h},w-w_{h})\lesssim\left\Vert u-u_{h}\right\Vert \left\Vert w-w_{h}\right\Vert ,
\]
where $\left\Vert \cdot\right\Vert $ is the $H^{1}(\Omega)$ norm
of the computational domain $\Omega$. For the subequations of the
PNPS problem, which are elliptic and whose primal and dual right hand
sides are continuous functionals on $H^{1}$, the \emph{a priori }estimate
we expect to hold is $\left\Vert u-u_{h}\right\Vert =\left\Vert w-w_{h}\right\Vert =O(h^{1})$
locally in the mesh size $h$; hence the product is $O(h^{2})$. In
3D, the local mesh size $h$ scales like $N^{-1/3}$, where $N$ are
the degrees of freedom. In summary, we expect 
\[
F(u-u_{h})=O(h^{2})=O(N^{-2/3}).
\]
 Similarly, in 2D we derive $F(u-u_{h})=O(N^{-1})$.%
}.

Practically speaking, we see that both adaptive schemes make it possible
to drive the approximation error of our main quantities of interest
down to a few percent, in the full 3D problem, with a single CPU core
on a desktop machine as used for creating Figure \ref{fig:adap}.
In case the axisymmetric 2D formulation can be applied, even much
lower tolerances can be achieved more quickly. Regarding the choice
of adaptive algorithm, from our results we would recommend to stick
with the more expensive Algorithm \ref{alg:adaptive} (which is still
cheap compared to solving the PNPS system). This algorithm does surprisingly
well given that it only uses information from solutions to the linear
Poisson-Boltzmann equation, and despite inexact discretization of
the curved geometry. The cheaper variant (Algorithm \ref{alg:adaptive-cheap})
seems to capture the rough regions of interest during the first few
refinement steps but may not be valid in the asymptotic limit.

\subsection{Finite-sized vs.\ point-sized target molecule\label{sub:Explicit-vs.-implicit}}

\begin{figure}
\begin{centering}
\subfigure[Electric force]{\includegraphics[height=5cm]{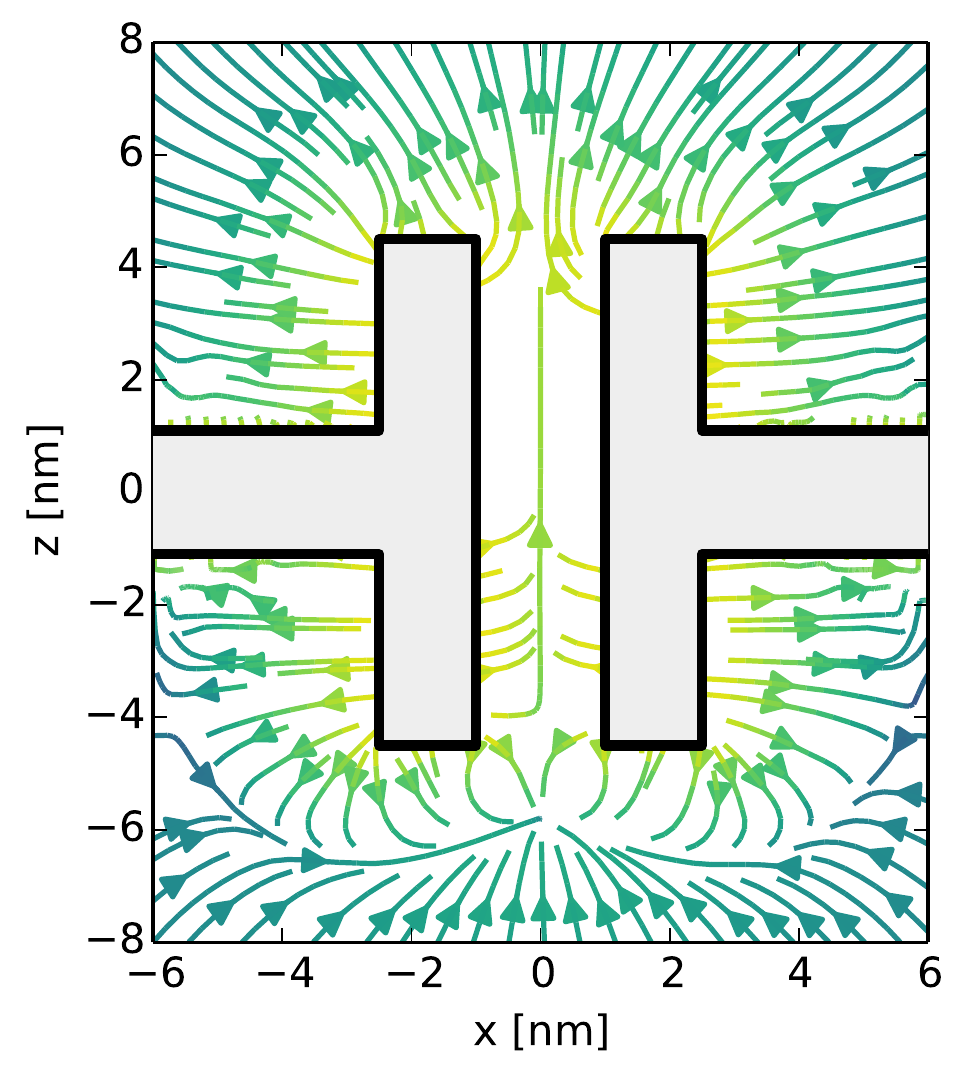}}\quad{}\subfigure[Drag force]{\includegraphics[height=5cm]{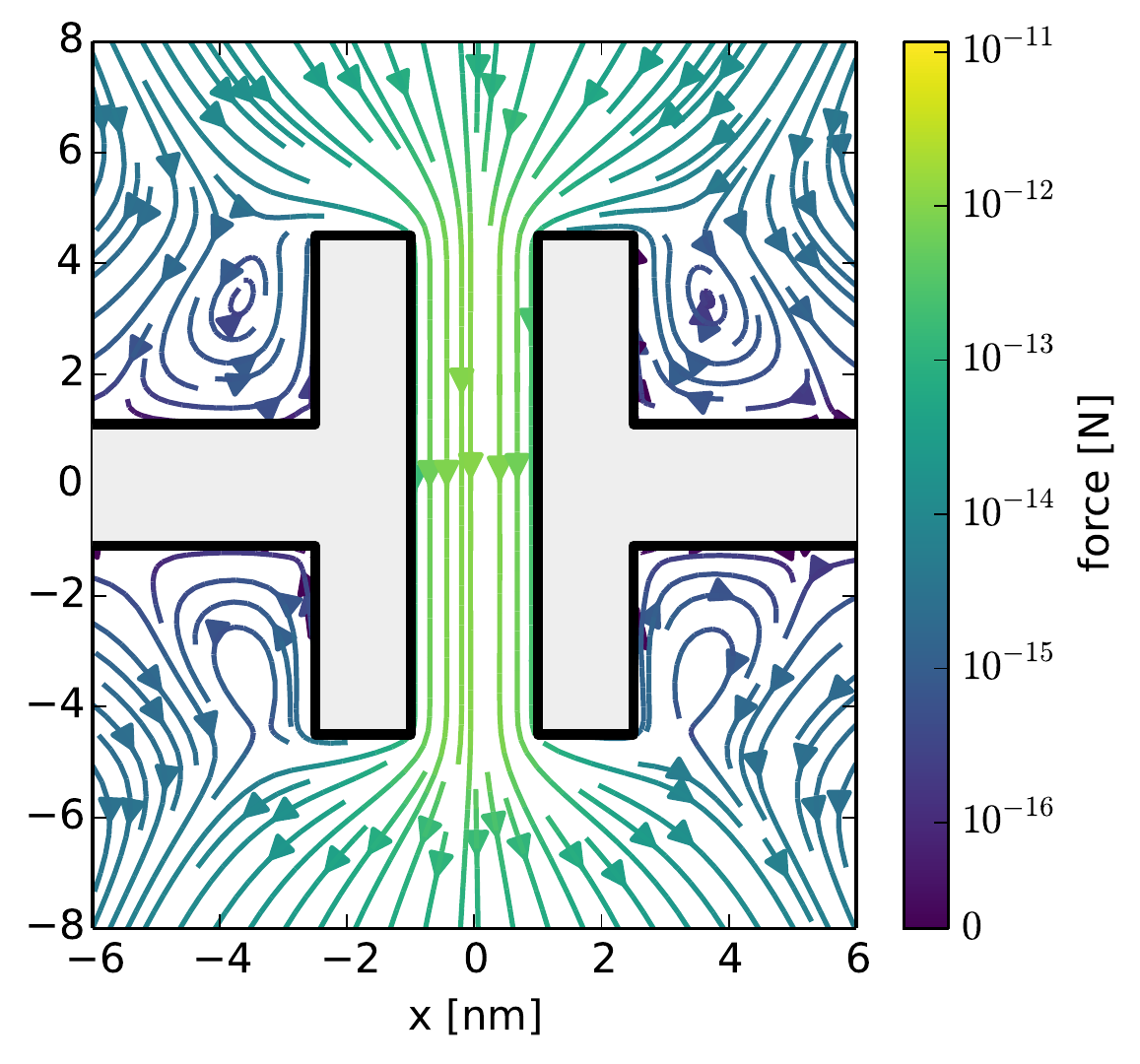}}
\par\end{centering}

\caption{\label{fig:Stream-plots}Streamline plots of PNPS force fields from
the point-sized molecule model. The molecule has a radius $r=0.5$
nm and a negative total charge of $-q$.}
\end{figure}

After having obtained a good picture of the numerical accuracy of
our adaptive PNPS solvers, we want to address the modeling question
put forward in Section \ref{sub:Physical-Quantities}: whether to
model the target molecule as \emph{finite-sized} and explicitly part
of the geometry or only implicitly (\emph{point-sized}), i.e., simulate
without molecule and only afterwards calculate the forces on a virtual
molecule from the electric and flow fields.

The point-sized molecule approach -- albeit less accurate -- is computationally
attractive, since the whole force field is obtained from a single
simulation, while in the finite-sized approach one has to compute
the force on a different mesh for every distinct molecule position.
In Figure \ref{fig:Stream-plots}, we show the electric and hydrodynamic
force fields obtained from the point-size model and an axisymmetric
simulation.

Qualitatively, the following physical phenomena can be observed in
Figure \ref{fig:Stream-plots}: As the molecule is negatively charged,
and the applied electric field is pointing downwards, the net electrical\emph{
}force $\F_{\text{el}}$ inside the pore is in the \emph{upper} direction
(Fig.~\ref{fig:Stream-plots}, left). On the other hand, the negative
surface charge of DNA (which makes up the pore wall) leads to crowding
of the pore by positive ions; the positive ions move downwards, causing
electroosmotic drag $\F_{\text{drag}}$ in the \emph{downward }direction
(Fig.~\ref{fig:Stream-plots}, right). From these two competing effects,
it is \emph{a priori} not clear whether the net force on the molecule
will be upwards or downwards. Outside the pore, the molecule is repelled
away by $\F_{\text{el}}$ due to the equally signed charges, which
leads to an energy barrier for entering the pore from either direction.

\begin{figure}
\begin{centering}
\begin{minipage}[t]{0.31\textwidth}%
\subfigure[Electric force along pore]{\resizebox{1.\textwidth}{!}{\includegraphics[width=1\paperwidth]{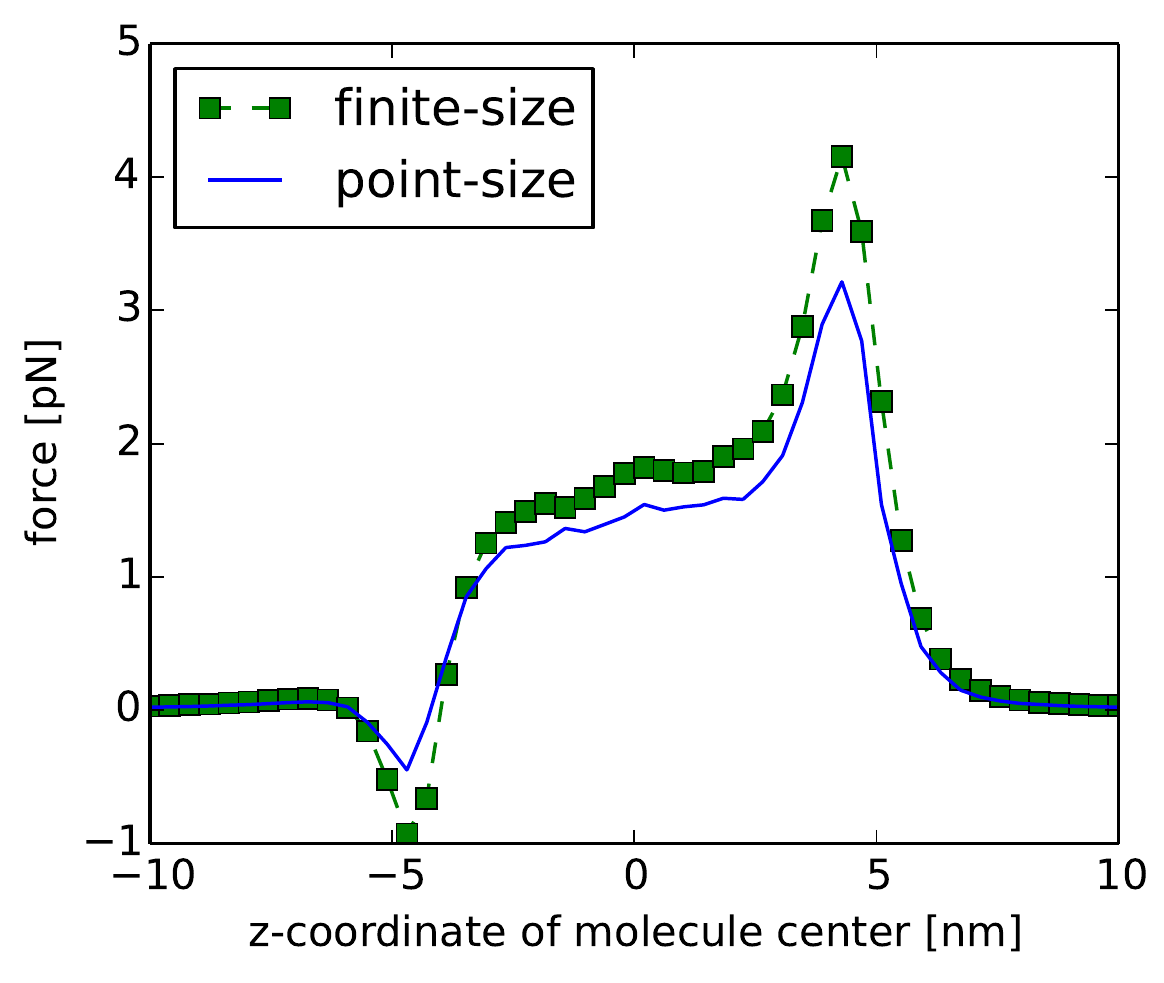}\label{fig:Fel}}}%
\end{minipage}\hspace*{\fill}%
\begin{minipage}[t]{0.31\textwidth}%
\subfigure[Drag force along pore]{\resizebox{1.\textwidth}{!}{\includegraphics[width=1\paperwidth]{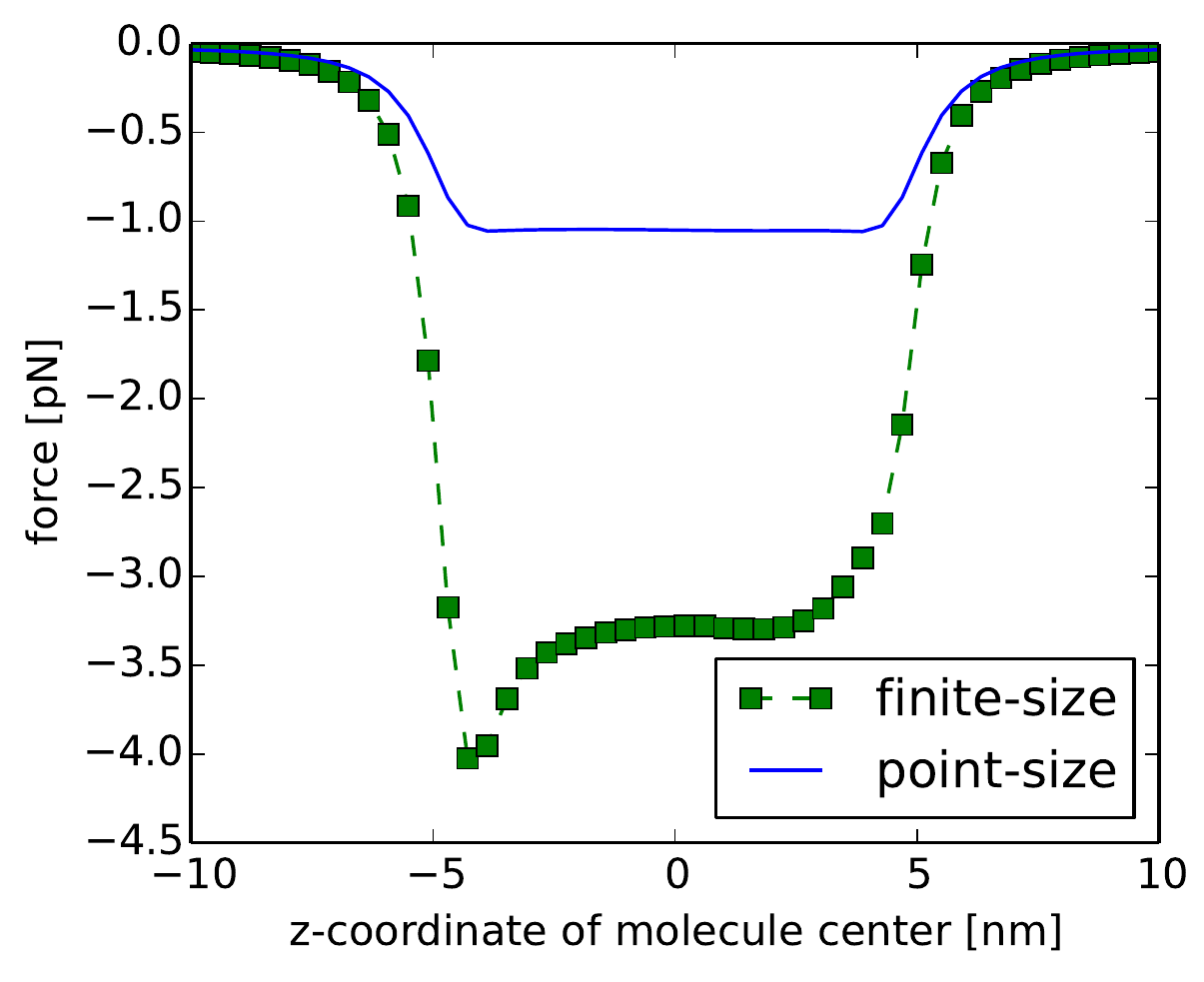}\label{fig:Fdrag}}}%
\end{minipage}\hspace*{\fill}%
\begin{minipage}[t]{0.31\textwidth}%
\subfigure[Total force along pore]{\resizebox{1.\textwidth}{!}{\includegraphics[width=1\paperwidth]{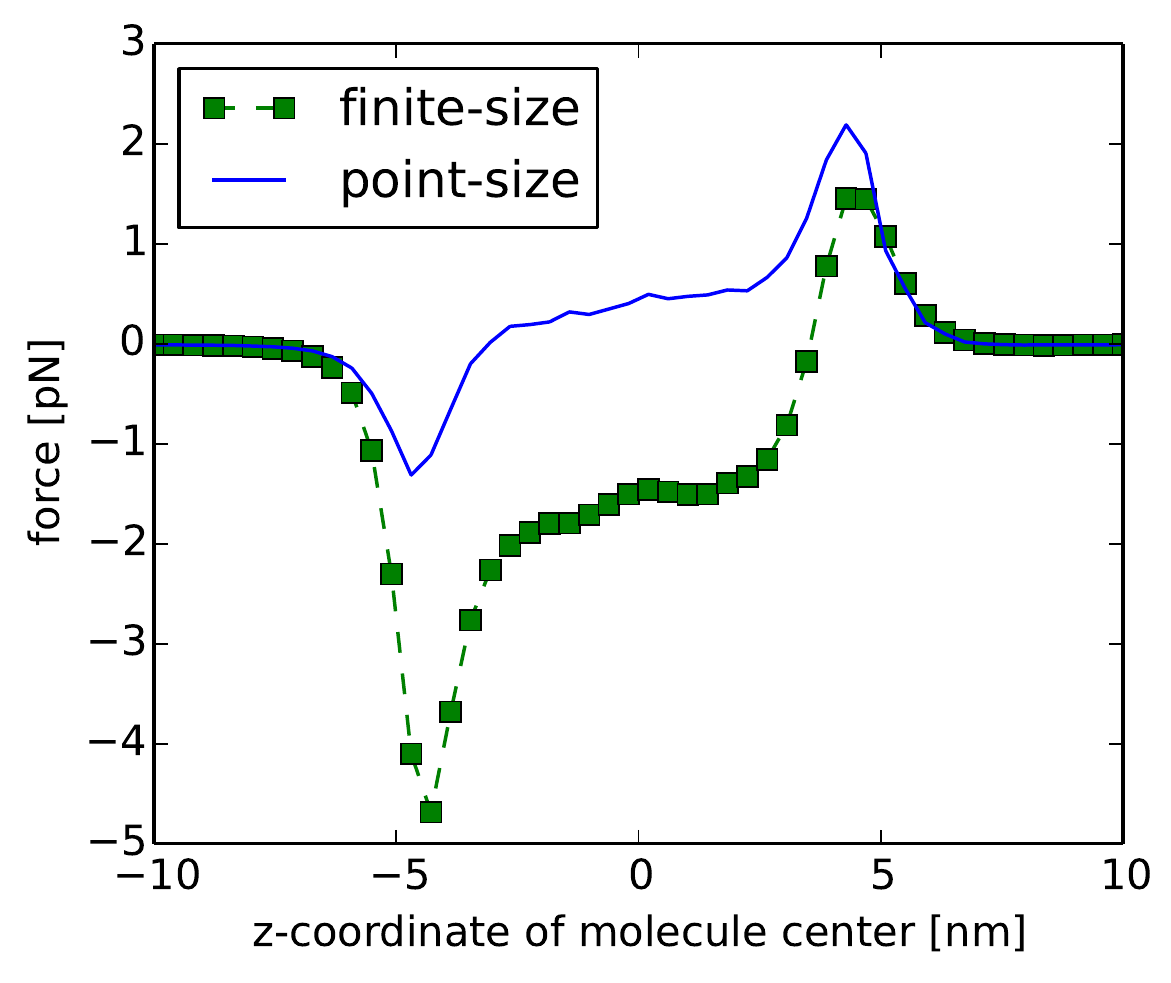}\label{fig:F}}}%
\end{minipage}
\par\end{centering}

\centering{}\caption{Force profiles along DNA nanopore, finite-size vs.\ point-size model.
The $z$-component of forces is plotted along the central pore axis,
where for the finite-size approach every data point corresponds to
a distinct geometry and thus simulation; for the point-sized approach
the whole curve is obtained from a single simulation. A negative total
force in some region means that the target molecule is, on average,
driven downward through the pore. In the point-size model, the magnitude
of drag force is underestimated to such an extent that the net total
force has a different direction than in the finite-size model. \label{fig:Forces}}
\end{figure}

\paragraph{Force profiles}

In Figure \ref{fig:Forces}, force profiles along the pore are compared
for the finite- and point-size formulation. For the point-sized molecule,
we calculate drag using Stokes' law $\F_{\text{drag}}^{*}(x)=6\pi\eta r\u(x)$
which is valid for a particle in bulk water. As we can see (Fig.~\ref{fig:Forces},
center), this leads to a strong underestimation of drag force inside
the pore. In the finite-size model, the drag force actually dominates
the electric force by a factor $2$, but is weakened by a factor $3$--$4$
in the point-size model; so the net total force comes out in the wrong
direction. For the electric force, agreement between both models looks
more reasonable, yet the point-size model fails to capture an additional
repelling force at the edges which can possibly be attributed to the
dielectric self-energy of the molecule~\citep{Corry2003}.

\paragraph{Improving the point-size model}

Looking at Figure \ref{fig:Forces}, it is clear that the point-size
model could be improved a lot by simply scaling the drag force by
a constant factor inside the pore. More generally, we propose to modify
the calculation of the force on a point-sized molecule by using
\[
\F^{*}(x)=\alpha(x)\F_{\text{el}}^{*}(x)+\beta(x)\F_{\text{drag}}^{*}(x).
\]
Here, $\alpha(x)$ and $\beta(x)$ are corrective factors yet to be
determined and $\F_{\text{el}}^{*}(x)$, $\F_{\text{drag}}^{*}(x)$
are the point-sized molecule forces given by \eqref{eq:Fel*} and
\eqref{eq:Fdrag*}. The corrections $\alpha$ and $\beta$ will be
obtained by some parametrization of the finite-size model, for instance
by evaluating the finite-size forces $\F_{\text{el}}(x)$ and $\F_{\text{drag}}(x)$
on a small number of points $x_{1},\ldots,x_{n}$ and constructing
$\alpha$ and $\beta$ by interpolation to satisfy
\[
\alpha(x_{i})=\frac{\F_{\text{el}}(x_{i})}{\F_{\text{el}}^{*}(x_{i})},\qquad\beta(x_{i})=\frac{\F_{\text{drag}}(x_{i})}{\F_{\text{drag}}^{*}(x_{i})},\qquad i=1,\ldots,n.
\]
In this way we ensure $\F^{*}(x_{i})=\F(x_{i})$ at the interpolation
points. The idea is that the expensive finite-size model has to be
evaluated only at a few points instead of hundreds or thousands, since
the rough overall shape of $\F^{*}$ is correctly determined by the
point-size model. Thus we hope to get the best of both models, efficiency
as well as accuracy.

\begin{figure}
\begin{centering}
\begin{minipage}[t]{0.31\textwidth}%
\subfigure[Correction of drag force  $\beta$(x) for different molecule charges]{\resizebox{1.\textwidth}{!}{\includegraphics[width=1\paperwidth]{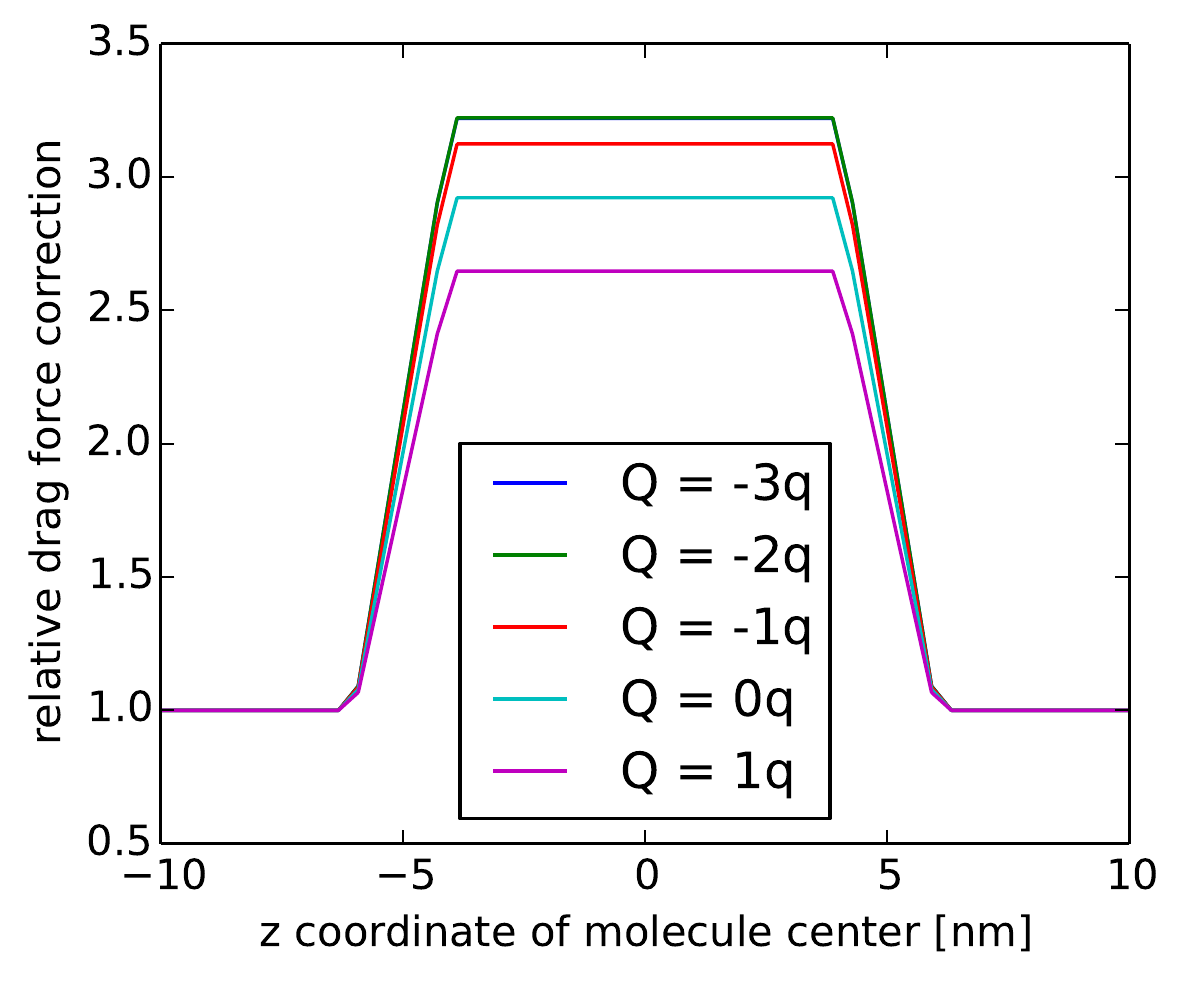}\label{fig:alpha}}}%
\end{minipage}\hspace*{\fill}%
\begin{minipage}[t]{0.31\textwidth}%
\subfigure[Total force, molecule charge $-q$]{\resizebox{1.\textwidth}{!}{\includegraphics[width=1\paperwidth]{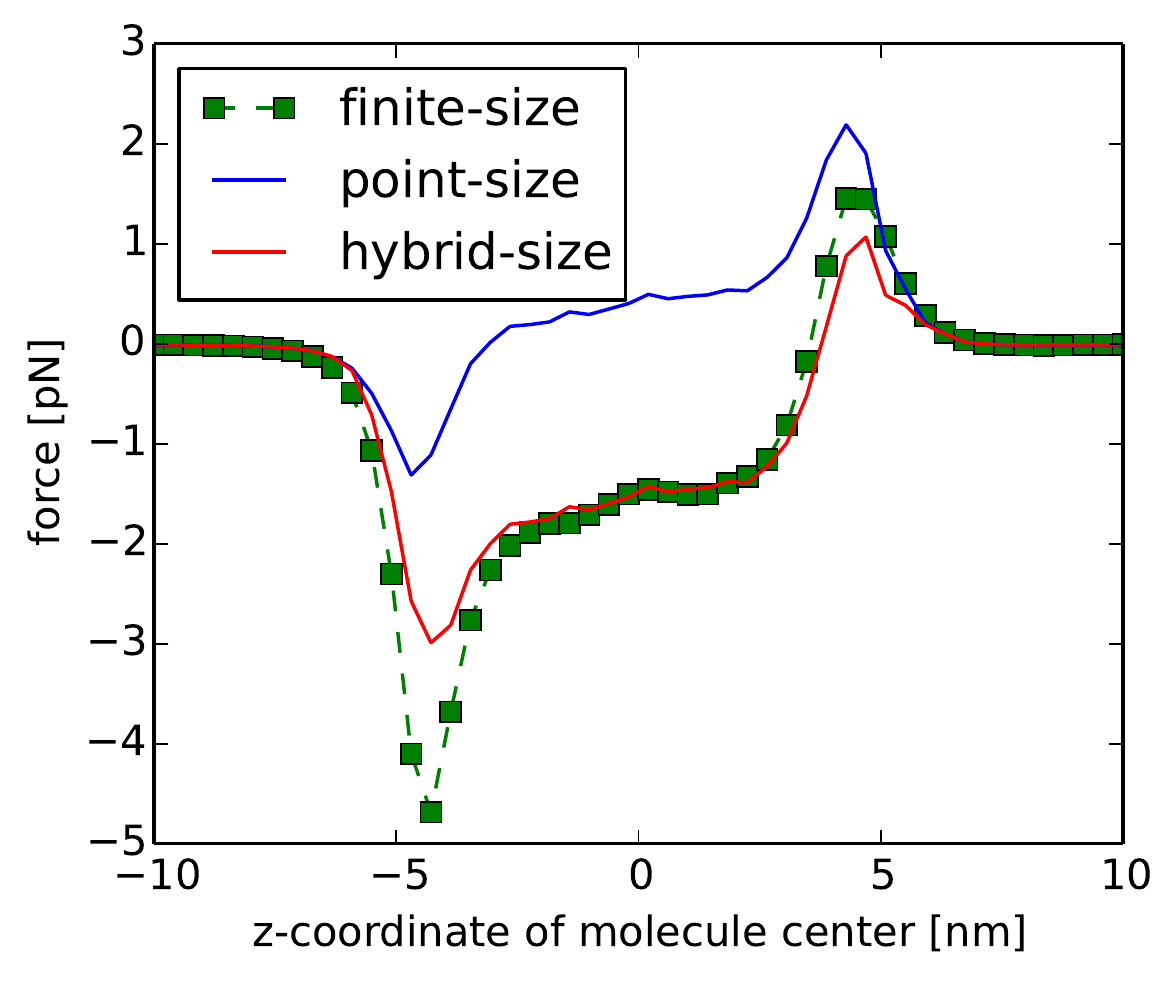}\label{fig:Fcorrect-1}}}%
\end{minipage}\hspace*{\fill}%
\begin{minipage}[t]{0.31\textwidth}%
\subfigure[Total force, molecule charge $+q$]{\resizebox{1.\textwidth}{!}{\includegraphics[width=1\paperwidth]{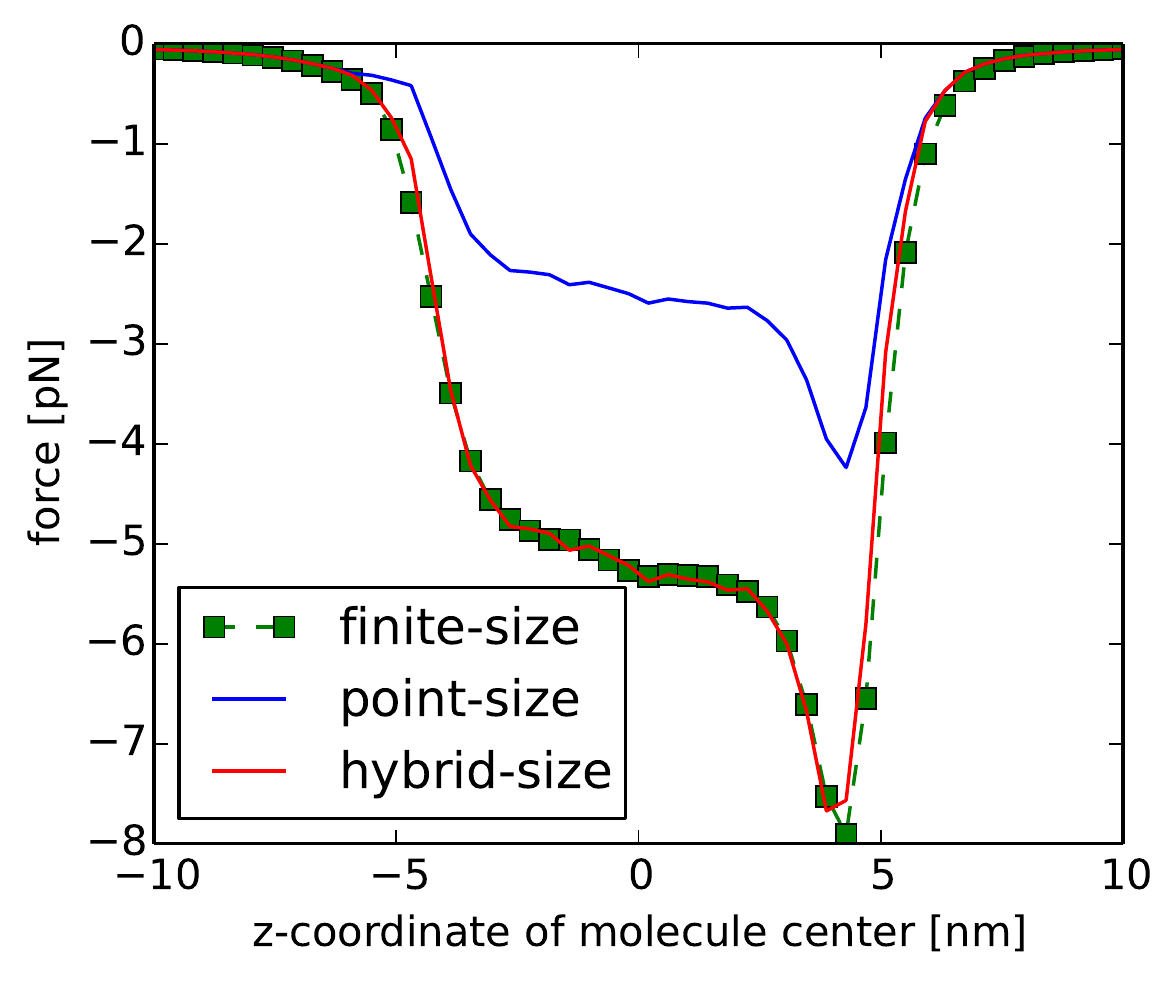}\label{fig:Fcorrect+1}}}%
\end{minipage}
\par\end{centering}

\centering{}\caption{Demonstration of the hybrid-size model for approximating the finite-size
model with minimal computational effort. Left: Correction factor computed
for the drag force, which measures by how much the finite-size model
is underestimated by the point-size model. Center/right: Corrected
force profiles for molecules with negative and positive unit charge,
respectively.\label{fig:drag-correction}}
\end{figure}

For demonstration, we apply this approach -- which we dub the \emph{hybrid-size
model} -- using only a \emph{single} interpolation point at the pore
center. The resulting correction factor $\beta(x)$ to the drag force
is shown in Figure \ref{fig:alpha}, for molecules of different charges.
We constructed $\beta$ so that is has the same (interpolated) value
throughout the pore; outside the pore, we set $\beta=1$ (assuming
that the point-size model is correct under bulk conditions), and in
the transition region near the pore edges, we interpolate linearly
between those two values. For $\alpha(x)$ the procedure is the same.

By using only one interpolation point, we deliberately ignore possible
edge effects captured by the finite- but not the point-size model.
Still, as shown in Figure \ref{fig:Fcorrect-1}, \subref{fig:Fcorrect+1},
this approach significantly improves on the pure point-size model
for molecules of both positive and negative charge, with only twice
the computational cost. This may be in part due to our specific nanopore
geometry, which has a constant radius along the entire pore; but the
approach is general enough to be applicable to irregularly shaped
pore proteins such as $\alpha$-hemolysin, by using more than one
interpolation point.

\paragraph{Comparison of molecule flux}

\begin{figure}
\begin{centering}
\begin{minipage}[t]{0.31\textwidth}%
\resizebox{1.\textwidth}{!}{\includegraphics[width=1\paperwidth]{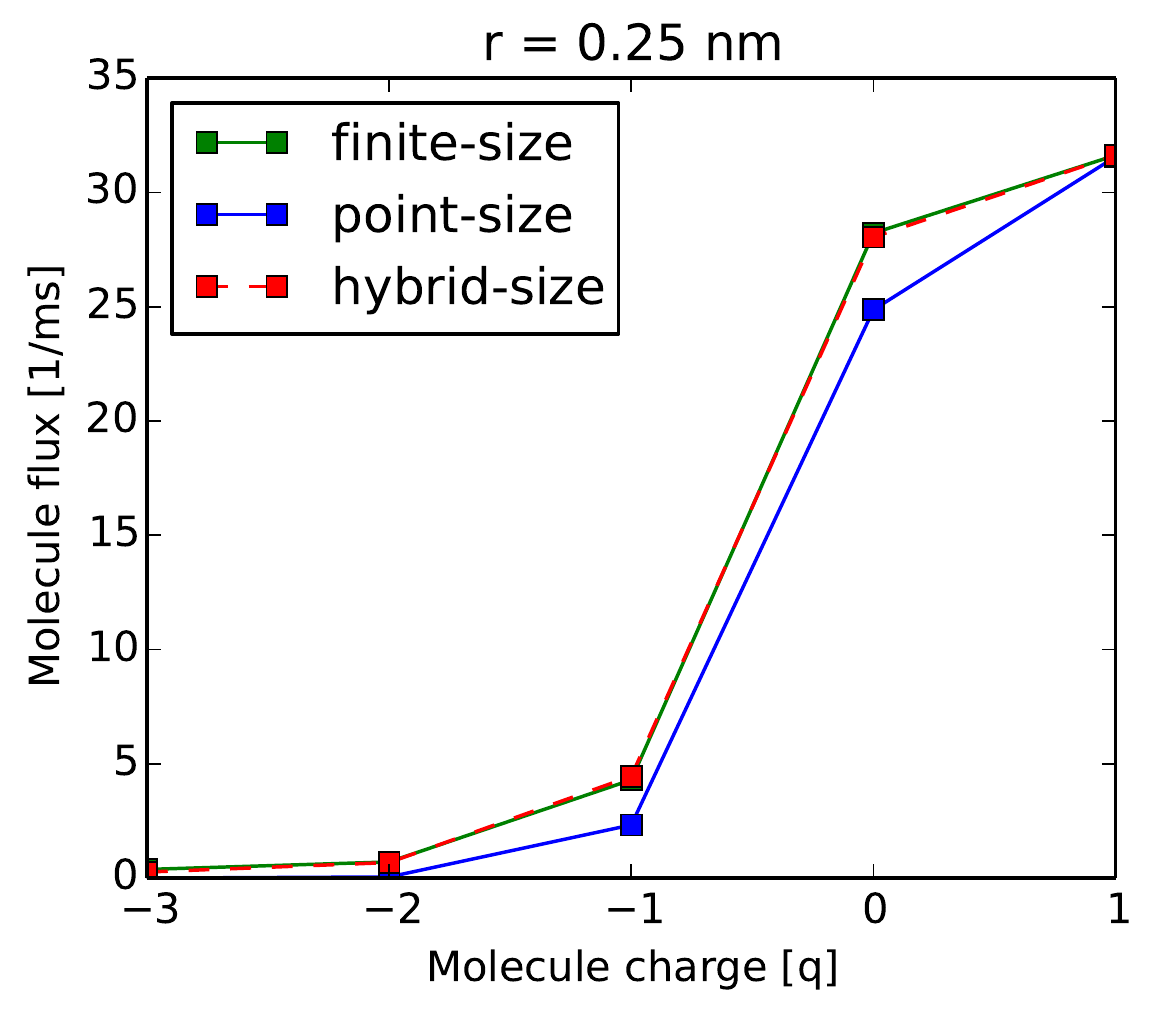}}%
\end{minipage}\hspace*{\fill}%
\begin{minipage}[t]{0.31\textwidth}%
\resizebox{1.\textwidth}{!}{\includegraphics[width=1\paperwidth]{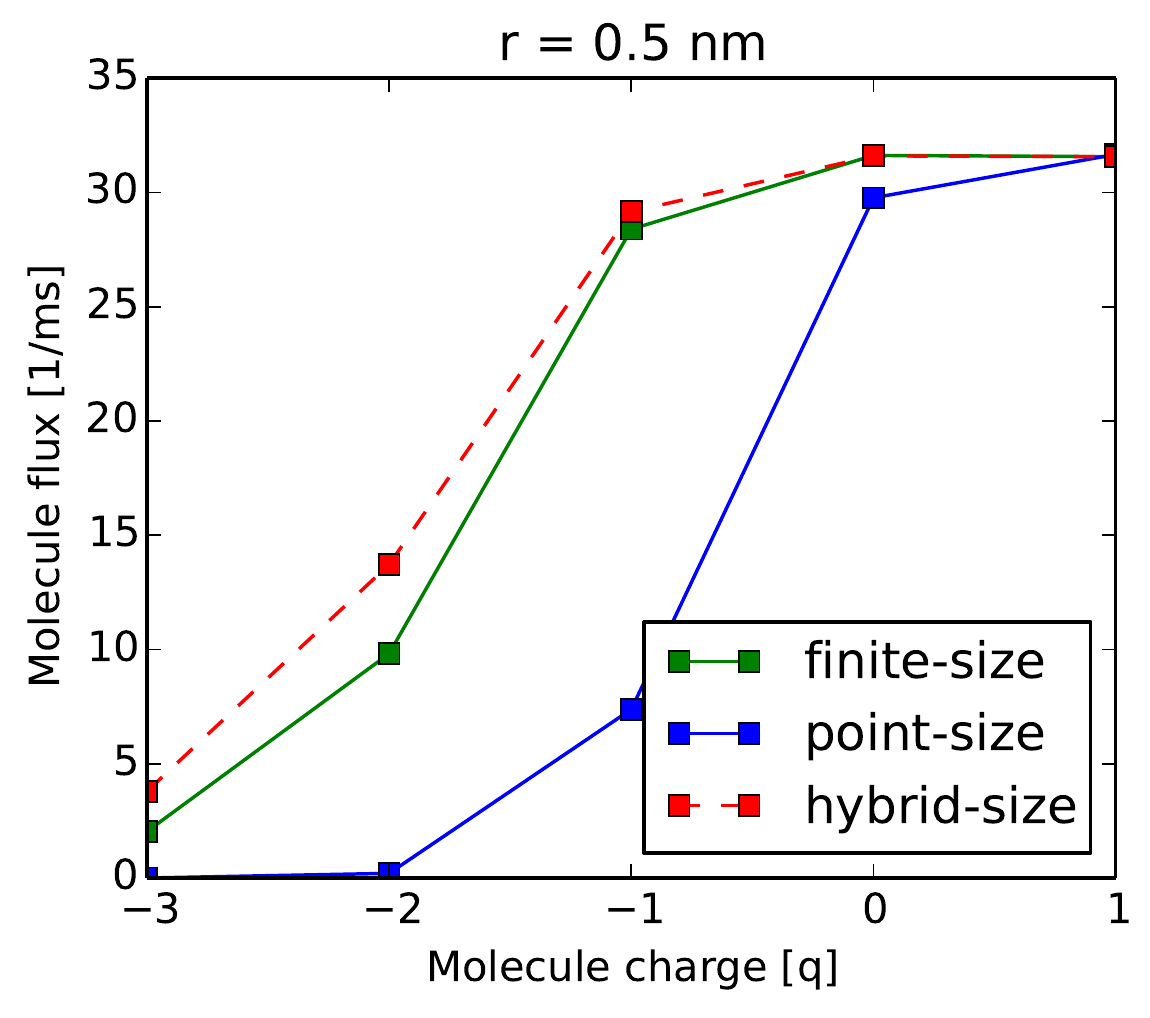}}%
\end{minipage}\hspace*{\fill}%
\begin{minipage}[t]{0.31\textwidth}%
\resizebox{1.\textwidth}{!}{\includegraphics[width=1\paperwidth]{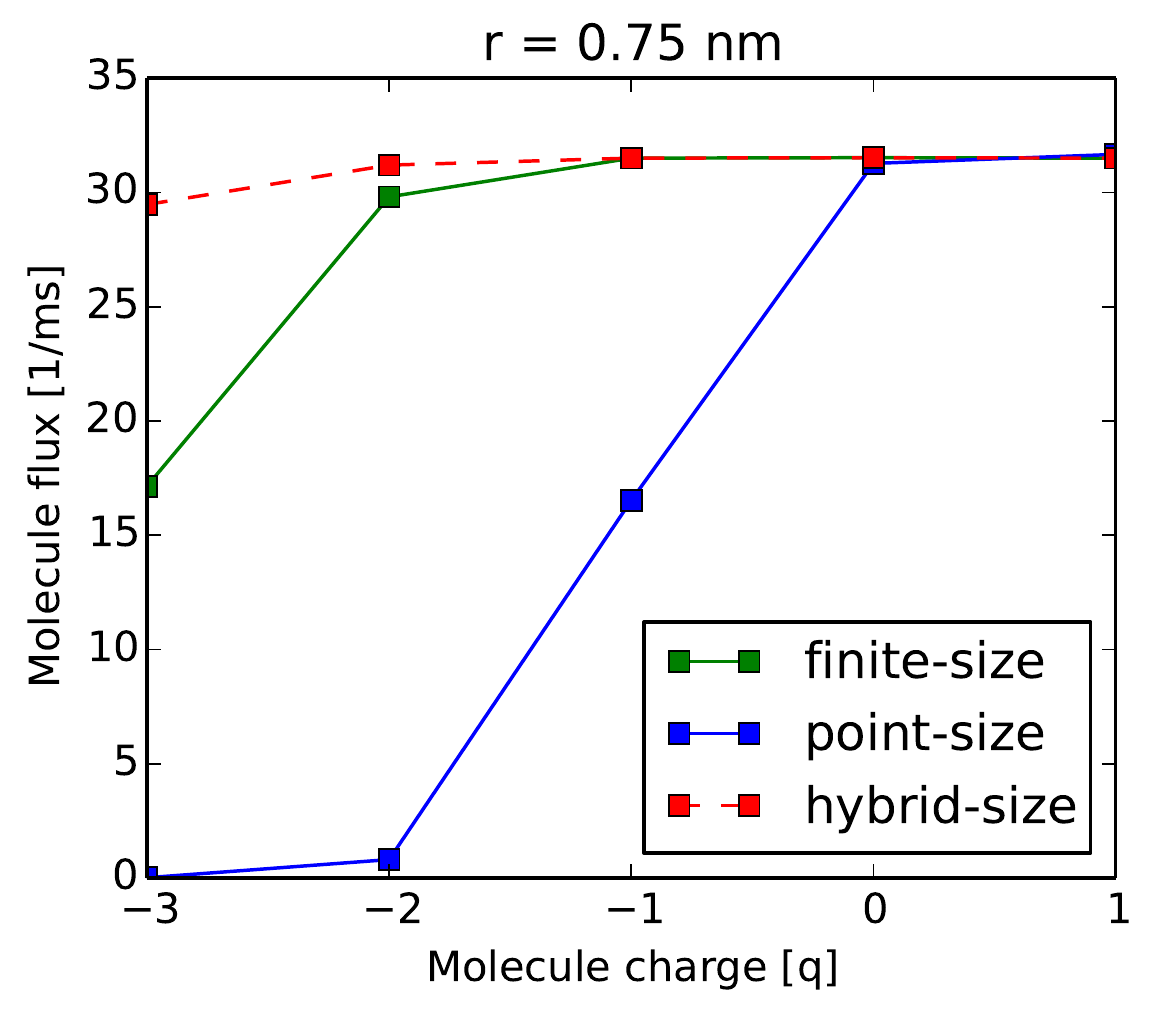}}%
\end{minipage}
\par\end{centering}

\centering{}\caption{Molecule flux after $1\mu{\rm s}$ for different molecule charges
and sizes and different models of the force. From left to right, the
radii are $r=0.25$, $0.5$, $0.75$ nm. The point-size model does
not match the results of the (expensive and more accurate) finite-size
model. For most setups however, the (cheap) hybrid-size model comes
relatively close.\label{fig:molecule-current}}
\end{figure}

Finally, we want to provide a more succinct way of relating our three
models (finite-\emph{, }point- and hybrid-size). In the end, we are
interested in how the PNPS force influences selectivity between different
molecule species. Therefore, the single number that represents each
model best is the flux of target molecules, i.e., the number of molecules
translocating the pore in a given time interval. We model transport
along the $z$-direction by a 1D diffusion equation $\frac{dc}{dt}=\frac{d}{dz}(-D\frac{dc}{dz}+\frac{D}{kT}Fc)$;
the molecules start out in the upper reservoir and diffuse through
the force field in the pore region. Details on the equation and how
we solve it are given in Appendix A.3. The term inside the brackets
$(-D\frac{dc}{dz}+\frac{D}{kT}Fc)$ is the flux density; multiplied
by the crosssectional area of the pore, it gives the total molecule
flux through the pore. We have computed the flux after a time of $1\mu{\rm s}$
for many different molecule charges and sizes, with results shown
in Figure \ref{fig:molecule-current}.

In Figure \ref{fig:molecule-current}, looking at molecule fluxes
from the finite-size model (green), we see that our sensor exhibits
non-trivial selectivity between molecules of different charges. Strongly
negatively charged molecules struggle to make their way through the
pore against the direction of voltage bias. This is more pronounced
for smaller molecules; for large molecules of radius $r=0.75$ nm,
there is almost no differentiation, as the charge-agnostic drag force
dominates. The latter effect is captured well by the hybrid-size model,
but not by the uncorrected point-size model. However, the hybrid-size
model as implemented with a single interpolation point underestimates
the energy barrier at the pore entrance for large molecules of charge
$-3q$, and therefore overstimates the flux there (Fig.~\ref{fig:molecule-current},
right). This effect vanishes for larger timescales, which is why we
stopped after $1\ \mu{\rm s}$ to reveal the limitations of our proposed
hybrid-size model.

\section{Conclusion\label{sec:Conclusion}}

We have implemented a 2D/3D finite element solver for the steady-state
Poisson-Nernst-Planck-Stokes system for the simulation of nanopore
sensors. To solve the nonlinear equations, we proposed three different
schemes for linearization in an attempt to clarify the amount of system
segregation that leads to the most efficient and robust solver. Two
methods emerged as roughly equal candidates, with the pure fixed-point
method being faster for small voltages and the hybrid method, if initialized
from the Poisson-Boltzmann equation, better suited to work across
a large range of voltages. For the fixed-point method, a new correction
to the Poisson equation was proposed that prevents blow-up and --
for small voltages -- enables convergence in less than $10$ iterations
without the need for damping.

The goal-adaptive mesh refinement schemes we proposed can efficiently
allocate computational resources towards the prediction of output
functionals. This proved indispensable to evaluate the electrophoretic
force on particles in 3D with accuracies of over $99\%$ on non-HPC
architectures. We experimented with a cheaper alternative to the goal-oriented
error estimation procedure established in \citep{bangerth2013adaptive}
by skipping the extrapolation step, but conclude that the extrapolation-based
algorithm remains the more robust choice. Interestingly, we have shown
that adaptivity can work in practice even if a radically simplified
model is used to compute the error indicators. This gives a total
speedup of roughly $50\%$, because we essentially save the effort
to compute all the solutions on coarser meshes that are not needed
in the end. The idea should be applicable to a variety of computational
problems where model simplifications are available.

For the modeling of selectivity in nanopore sensors, we find that
assuming a point-sized target molecule, although computationally attractive,
severely reduces the predictiveness of our simulations. For practical
applications, it is therefore recommended to model the molecule explicitly
and calculate the force for each possible molecule position separately.
However, if this is too expensive, a carefully calibrated hybrid model
that we proposed may also capture most features of the force field
and give results similar to the full finite-size model.

\section*{Acknowledgements}

We acknowledge support through the Austrian Science Fund (FWF) START
Project No. Y660 \emph{PDE Models for Nanotechnology}.


\bibliographystyle{elsarticle-harv}
\bibliography{friends,numerics,nanopores}

\section*{Appendix }

\subsection*{A.1 Semi-analytical test problem \label{sub:(Almost)-analytical-test}}

To validate our 2D and 3D solvers, we look at a simple model problem
where the whole PNPS system can be reduced analytically to the solution
of a scalar 1D boundary value problem. Since the 1D problem can be
solved to high accuracy with negligible effort, it serves our purpose
just as well as an exact analytical solution would. Unlike manifactured
test problems often employed in academia, we do not introduce artificial
right hand side terms. The problem actually describes, albeit simplified,
a physically relevant situation: that of a nanopore where the pore
length is large compared to the radius and the applied voltage is
sufficiently low to have negligible influence on the ion distribution.

The domain is a plain cylindrical tube of radius $R=1$ nm and length
$L=4$ nm and is meant to represent an inner part of the nanopore
(far away from the pore entrances on either side); the only dielectric
material is water. Using axial symmetry, we write the PNPS solutions
as functions of the radial component $r$ and the height $z$. The
exact potential is prescribed as 
\[
\phi(r,z)=U_{T}\phi^{*}(r)-E_{0}z,
\]
where $E_{0}$ is a given constant external field strength, $U_{T}=kTq^{-1}$
is the thermal voltage and the dimensionless potential $\phi^{*}(r)$
solves a 1D radial Poisson-Boltzmann equation
\[
\frac{1}{r}\frac{d}{dr}\left(r\frac{d\phi^{*}}{dr}\right)=\frac{1}{\lambda_{D}^{2}}\sinh(\phi^{*}).
\]
Here $\lambda_{D}=\sqrt{\frac{\varepsilon kT}{2Fc_{0}q}}$ is the
Debye length; the potential is subject to the boundary condition $U_{T}\frac{d\phi^{*}}{dr}(R)=\rho$
for a given constant surface charge density $\rho$. This equation
is readily solved e.g.\ by 1D FEM.

The other unknowns depend only on $r$ and can all be obtained from
$\phi^{*}(r)$: Ion concentrations are given by the Boltzmann factors
\[
c^{\pm}(r)=c_{0}\exp\left(\mp\phi^{*}(r)\right)
\]
for any given bulk concentration $c_{0}$; the velocity is of the
form $\boldsymbol{u}=(0,0,u)^{T}$ with
\[
u(r)=\frac{\varepsilon E_{0}}{\eta}U_{T}\left[\phi^{*}(r)-\phi^{*}(R)\right];
\]
and the pressure is
\[
p(r)=-2Fc_{0}U_{T}\cosh(\phi^{*}(r))+p_{0}
\]
with an arbitrary constant $p_{0}$, which we fix by requiring $p(0)=0$.
A mathematical derivation of this exact solution can be found in Appendix
A.2.

\paragraph*{Numerical results}

We solve the problem both in 3D and in axisymmetric 2D variables.
Dirichlet or Neumann boundary conditions are applied to match the
exact solutions; we have tried to encode as little information about
the exact solutions as possible into the boundary conditions. For
instance, for the Stokes equation it suffices to enforce the usual
no-slip condition $\u=0$ on the wall, the Neumann-type condition
$n\cdot\grad\u=0$ at the upper and lower ends, and the no pressure
condition $p=0$ somewhere in the center. The numerical values used
for the free parameters were $E_{0}=-100$ mV, $\rho=-0.05$ $\text{C}/\text{m}^{2}$
and $c_{0}=300$ $\text{mol}/\text{m}^{3}$, respectively. The Debye
length, which determines the width of the double layer and thus the
multiscale nature of the problem, is $\lambda_{D}=0.56$ nm in this
case. Comparing this with the pore radius of $R=1$ nm, we see that
we have posed ourselves a rather easy problem numerically (in the
sense that a coarse mesh will suffice to resolve all features). Our
test problem can be turned into a challenging benchmark problem by
increasing the pore radius such that $R/\lambda_{D}$ is large.

\begin{figure}
\begin{centering}
\begin{minipage}[t]{0.42\textwidth}%
\subfigure[Ion concentrations]{\includegraphics[height=4cm]{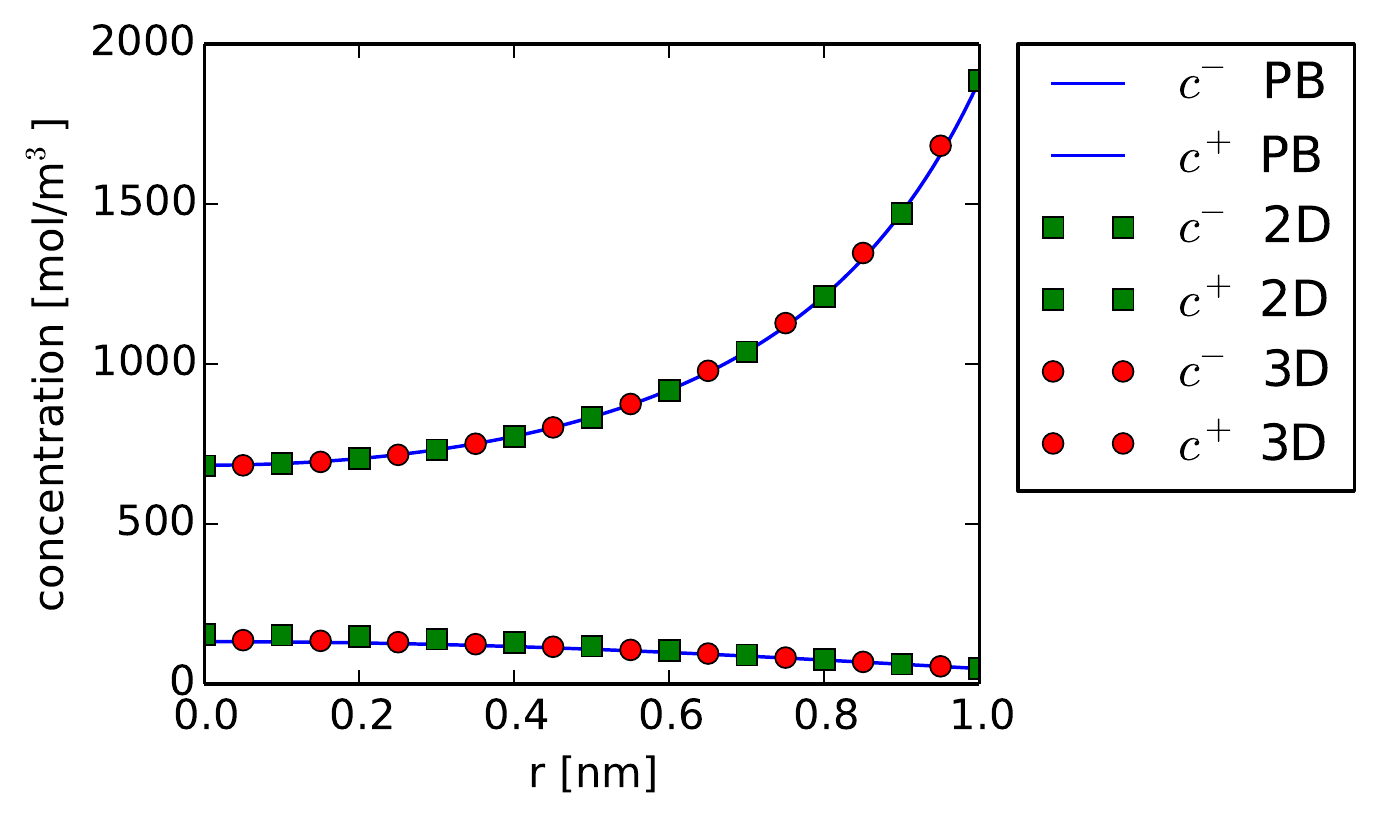}\label{fig:analytical-cross-c}}%
\end{minipage}\hspace*{\fill}%
\begin{minipage}[t]{0.42\textwidth}%
\subfigure[Velocity profile]{\includegraphics[height=4cm]{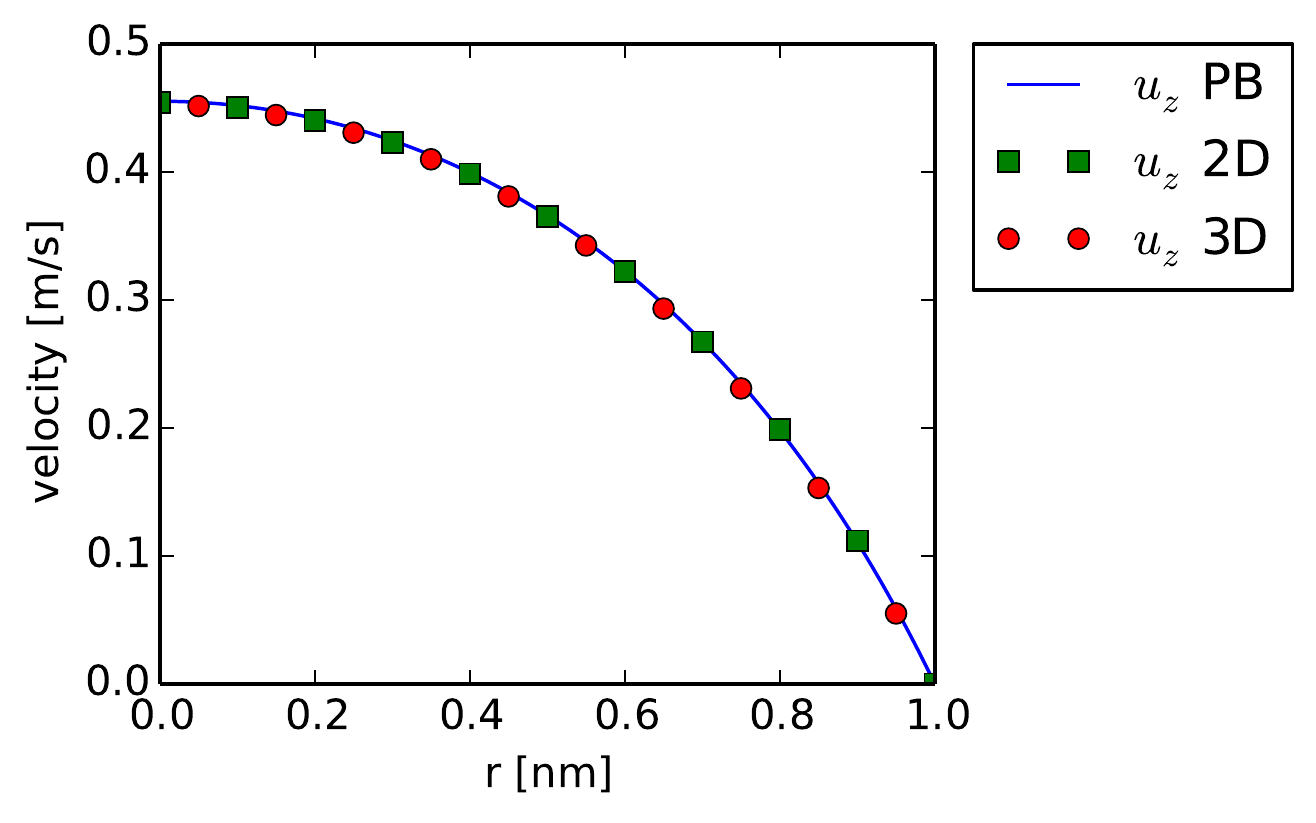}\label{fig:analytical-cross-u}}%
\end{minipage}
\par\end{centering}

\centering{}\caption{Cross-sectional plots for the analytical test problem with three different
numerical methods: the 1D PB equation and the 2D and 3D PNPS solvers.
\label{fig:analytical-cross}}
\end{figure}

In Fig.~\ref{fig:analytical-cross} we compare cross-sectional plots
of ion concentrations and the velocity profile coming from the three
different numerical formulations: the 1D Poisson-Boltzmann equation
-- which we take as the \emph{exact} \emph{solution} --, the axisymmetric
2D model and the full 3D PNPS system. We used a uniform mesh width
of $h=0.1$ nm in these calculations, which corresponds to about $1$k
mesh elements in 2D and $50$k in 3D. It is reassuring that the three
results in Fig.~\ref{fig:analytical-cross} are almost indistinguishable.

\begin{figure}
\begin{centering}
\begin{minipage}[t]{0.48\textwidth}%
\subfigure[Convergence with nonlinear iteration]{\includegraphics[height=4cm]{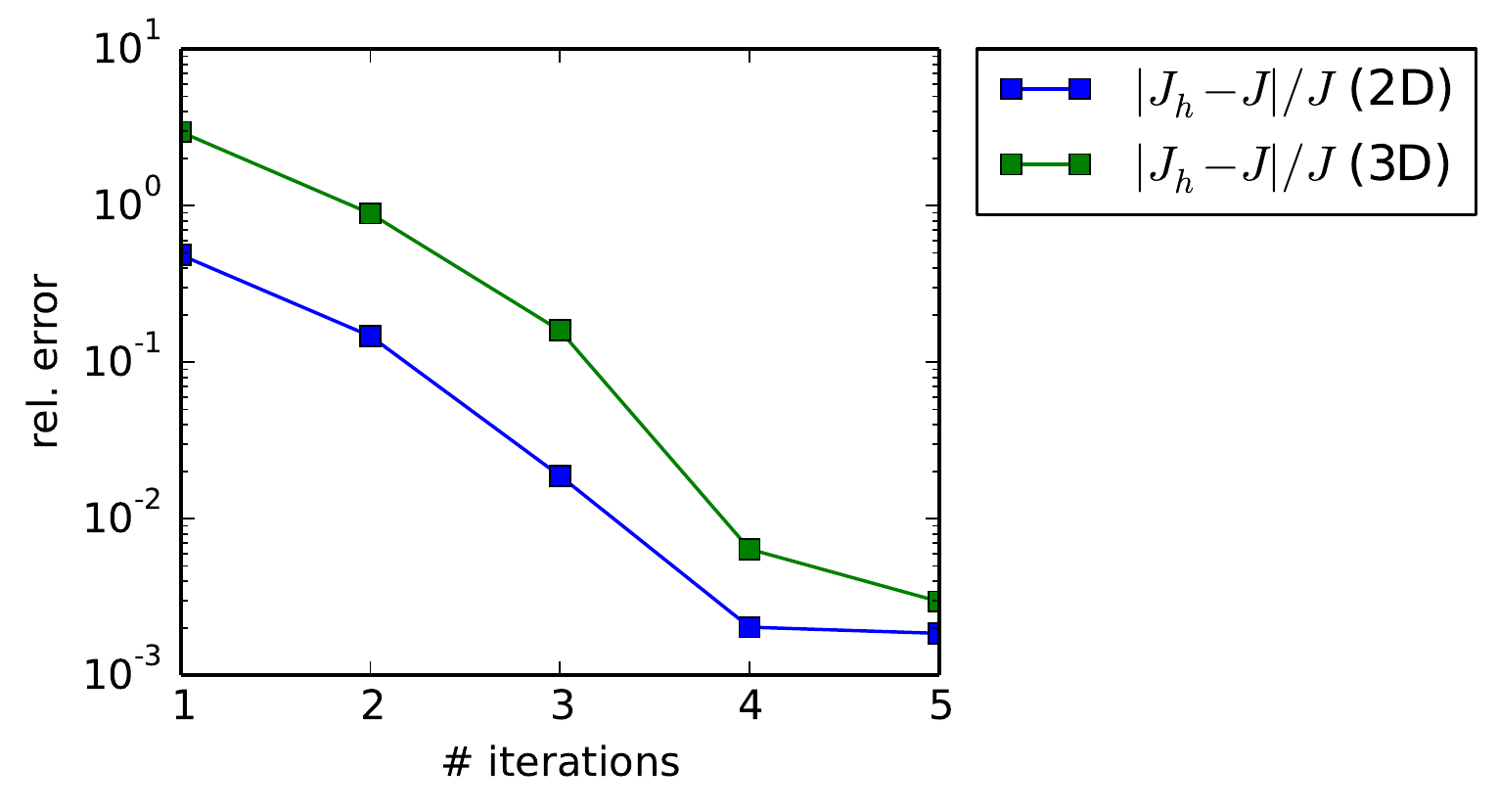}\label{fig:ana-hybrid-conv}}%
\end{minipage}\hspace*{\fill}%
\begin{minipage}[t]{0.48\textwidth}%
\subfigure[Convergence with mesh refinement]{\includegraphics[height=4cm]{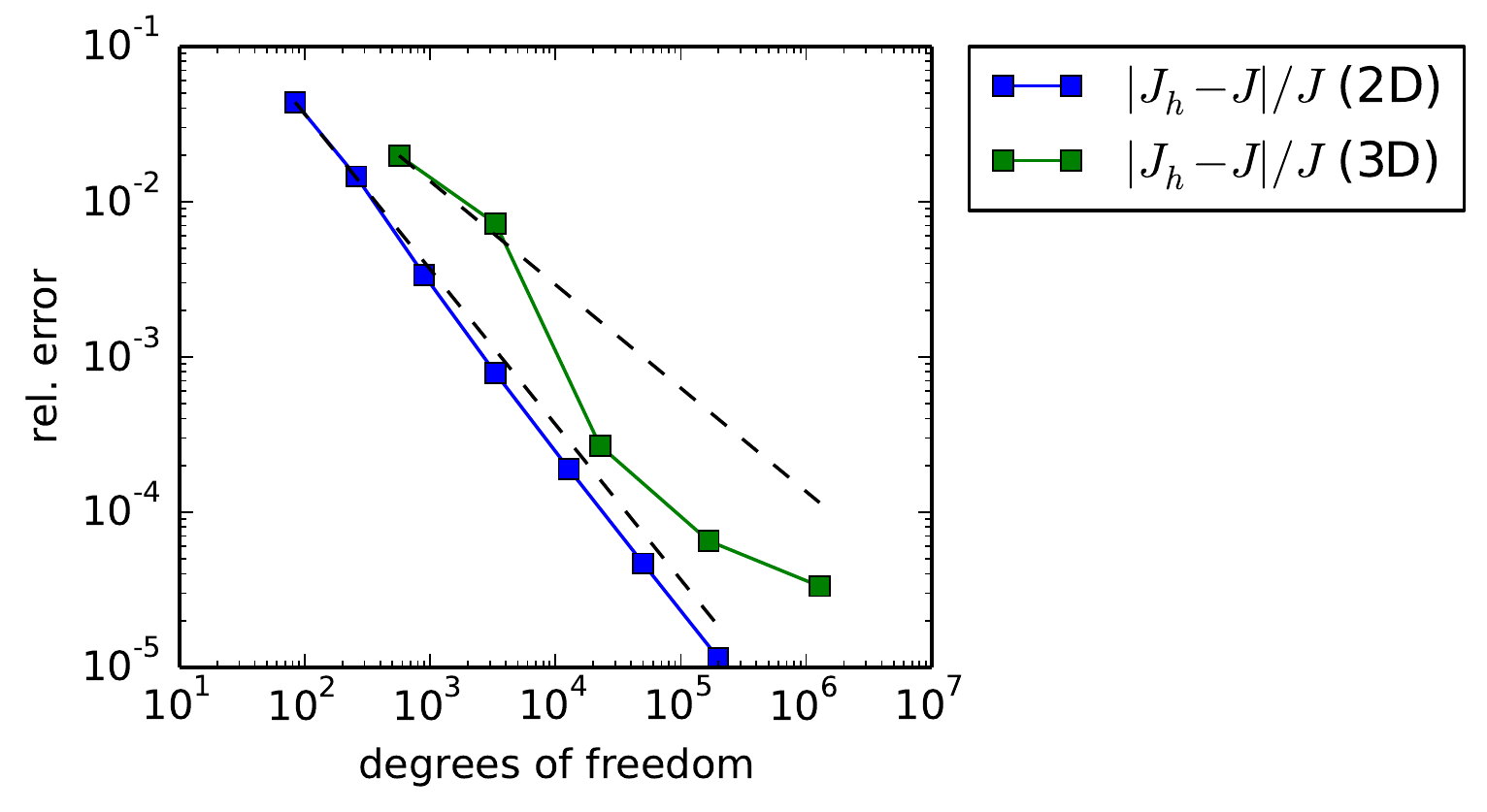}\label{fig:ana-adapt-conv}}%
\end{minipage}
\par\end{centering}

\centering{}\caption{Convergence to exact solution. The relative error is defined as $|J_{h}-J|/|J|$
where $J$ is the exact current from the 1D model and $J_{h}$ is
the output of the PNPS solvers. \subref{fig:ana-hybrid-conv} Error
against number of iterations in the hybrid linearization scheme. \subref{fig:ana-adapt-conv}
Error against degrees of freedom $N$ for uniform mesh refinement.
The theoretical $N$ dependency is $O(N^{-1})$ in 2D and $O(N^{-2/3})$
in 3D, indicated by dashed lines. \label{fig:analytical-convergence}}
\end{figure}

For a quantitative assessment of the solver accuracy, we calculate
the total ion current through the pore, defined as 
\[
J=F\int\left(j_{z}^{+}-j_{z}^{-}\right),
\]
where $j_{z}^{\pm}$ is the $z$ component of the ion flux defined
in \eqref{eq:nernst-planck}; the integral is taken over any horizontal
cross-section of the pore. In practice, it is more robust to average
over many crosssections by taking a volume integral, which is what
we do. The ion current involves the potential, both ion concentrations
and the velocity and therefore provides a suitable measure of accuracy.
In Fig.~\ref{fig:ana-hybrid-conv} we show the relative error of
$J$ compared to the exact solution against number of iterations of
our hybrid Newton/fixed-point scheme, with fixed mesh size $h=0.2$
nm. After about five iterations a converged state is reached where
the error stops changing; at this point, the linearization error is
dominated by the discretization error.

To study the discretization error, in Fig.~\ref{fig:ana-adapt-conv}
the mesh is refined uniformly starting from a coarse initial mesh,
and the resulting error curve compared to the number of degrees of
freedom is shown. Note that every data point in the latter figure
corresponds to the final converged state of Fig.~\ref{fig:ana-hybrid-conv}.
Different asymptotic slopes for 2D and 3D arise as derived in Section
\ref{sub:Assessment-of-goal-adaptive}, confirming that the error
in $J$ is $O(h^{2})$. This provides a strong validation for the
correctness of our implementation. More generally speaking, the results
also indicate that for the current we should attain high numerical
accuracy quite easily in simulations similar to this test problem,
without any goal-specific adaptivity.

\subsection*{A.2 Derivation of test problem\label{sub:derivation-analytical-test}}

First, assume that the potential $\phi(r,z)$ is a superposition of
a part resulting from a given constant external field $E_{0}$ in
the $z$-direction and a part which depends only on the radius. Thus,
\[
\phi(r,z)=\psi(r)-E_{0}z.
\]
Second, the ion concentrations are assumed to depend only on $r$
and to be given by Boltzmann factors
\[
c^{\pm}(r)=c_{0}\exp\left(\mp\frac{q\psi(r)}{kT}\right).
\]
Third, for the Stokes equation we assume the velocity is of the form
$\u=(0,0,u_{z})^{T}$ with $u_{z}=u_{z}(r)$ and pressure $p=p(r)$,
i.e., the velocity field points only in the $z$-direction but is
constant in this direction.

These three assumptions, which can be encoded in the boundary conditions,
determine a solution to the PNPS system as follows. Since the second
derivatives of the external part of the potential $-E_{0}z$ vanish
and because of the assumed form of the ion concentrations, the Poisson
equation reduces to a 1D Poisson-Boltzmann equation in radial direction
\[
\varepsilon\frac{1}{r}\partial_{r}(r\partial_{r}\psi)=2Fc_{0}\sinh\left(\frac{q\psi}{kT}\right)
\]
with Neumann boundary condition $\partial_{r}\psi(R)=\rho$ for some
given (constant) surface charge density $\rho$. The permittivity
$\varepsilon$ is constant since our domain only consists of fluid.

The Stokes equation splits into a $x$, $y$ part only for the pressure
(since $u_{x}=u_{y}=0$) and a $z$ part only for the velocity (since
$\partial_{z}p=0$),
\begin{eqnarray}
\partial_{r}p & = & -F(c^{+}-c^{-})\partial_{r}\psi,\label{eq:analytical-p}\\
-\eta\frac{1}{r}\partial_{r}(r\partial_{r}u_{z}) & = & -F(c^{+}-c^{-})E_{0}.\label{eq:analytical-u}
\end{eqnarray}
Equation \eqref{eq:analytical-p} can be solved by noting that the
right hand side can be written as a derivative
\[
-F(c^{+}-c^{-})\partial_{r}\psi=-2Fc_{0}\sinh\left(\frac{q\psi}{kT}\right)\partial_{r}\psi=\partial_{r}\left[-2Fc_{0}\frac{kT}{q}\cosh\left(\frac{q\psi}{kT}\right)\right],
\]
which is equal to $\partial_{r}p$, thus yielding
\[
p(r)=-2Fc_{0}\frac{kT}{q}\cosh\left(\frac{q\psi(r)}{kT}\right)+p_{0}
\]
for some arbitrary constant $p_{0}$. For equation \eqref{eq:analytical-u},
we write on the other hand 
\[
-F(c^{+}-c^{-})=\varepsilon\frac{1}{r}\partial_{r}(r\partial_{r}\psi)
\]
and obtain
\[
\partial_{r}(r\partial_{r}u_{z})=-\frac{\varepsilon E_{0}}{\eta}\partial_{r}(r\partial_{r}\psi),
\]
which combined with the no-slip condition $u(R)=0$ yields the solution
\[
u(r)=\frac{\varepsilon E_{0}}{\eta}\left[\psi(R)-\psi(r)\right].
\]
Finally, the form of these solutions can be simplified by writing
the potential as $\psi(r)=\frac{kT}{q}\phi^{*}(r)$, where $\phi^{*}(r)$
is dimensionless.

\subsection*{A.3 Diffusion equation for molecule flux\label{sub:A.3-1D-Diffusion}}

We show how to compute the flux of target molecules for the results
in Figure \ref{fig:molecule-current}, Section \ref{sub:Explicit-vs.-implicit}.
For simplicity, we model flux only in the $z$-direction. Given a
force profile $F(z)$ which is obtained from the PNPS model (e.g.\ Fig.~\ref{fig:Forces}),
the time-dependent diffusion equation for target molecules reads\begin{subequations}\label{eq:1D-diffusion}

\begin{eqnarray}
j & = & -D\frac{dc}{dz}+\frac{D}{kT}Fc,\\
\frac{dj}{dz} & = & \frac{dc}{dt}.
\end{eqnarray}
\end{subequations}Here, $c(z,t)$ is the molecule concentration;
$j(z,t)$ is the molecule flux density; and $D(z)$ is the position-dependent
diffusion coefficient. We solve this equation on the symmetric interval
$|z|\le1\ \mu{\rm m}$, with $|z|\le4.5\ {\rm nm}$ defining the pore
region. For the initial condition, we use
\[
c(z,0)=\begin{cases}
c_{0}, & z>4.5\ \text{nm,}\\
0, & \text{elsewhere,}
\end{cases}
\]
with $c_{0}=1\text{ mol}/\text{m}^{3}$. Thus, the molecules start
out uniformly distributed in the upper reservoir, and will diffuse
downwards through the pore region. At the reservoir boundaries $z=\pm1\ \mu{\rm m}$,
Neumann conditions $\frac{dj}{dz}=0$ are applied, modeling hard walls.
The diffusion coefficient is modeled by a generalized Stokes-Einstein
relation 
\[
D(z)=\frac{kT}{6\pi\eta r}d(z)\qquad\text{where }d(z)=\begin{cases}
d_{r}<1 & \text{in the pore,}\\
1 & \text{in bulk.}
\end{cases}
\]
Here, $r$ denotes the molecule radius. The reduced relative diffusivity
$d_{r}$ inside the pore depends on $r$ and is taken from the tables
in \citep{Paine1975}.

We solve \eqref{eq:1D-diffusion} using backward Euler in time applied
to a $P^{1}$ finite element space discretization on a uniform grid.
For the results in Fig.~\ref{fig:molecule-current}, we used a grid
spacing of $h=0.2$ nm ($10$k nodes), a timestep of $dt=10$ ns and
stopped after $100$ steps or $t=1\ \mu{\rm s}$. The final-time downwards
flux density $-j(\cdot,t)$ was averaged over the pore region and
multiplied by the cross-sectional pore area $A=\pi(1{\rm nm})^{2}$
and the Avogrado constant $N_{A}$ to yield the molecule flux in numbers
per second.
\end{document}